\documentclass[prb,twocolumn,amsmath,amssymb,superscriptaddress]{revtex4-2}
\usepackage{epsfig,graphicx,graphics,float}
\usepackage[T1]{fontenc}
\usepackage[latin9]{inputenc}
\usepackage{appendix}
\usepackage{amscd}
\usepackage{bm}
\usepackage{psfrag} 
\usepackage{bbm} 
\usepackage{babel}
\usepackage{wasysym}
\usepackage{mathrsfs}
\usepackage{xcolor}
\usepackage{array}
\usepackage{subfigure}
\usepackage{verbatim} 
\usepackage{braket}
\usepackage{soul}
\usepackage[normalem]{ulem}

\newcommand{\be}{\begin{equation}}
\newcommand{\ee}{\end{equation}}
\newcommand{\bea}{\begin{eqnarray}}
\newcommand{\eea}{\end{eqnarray}}
\newcommand{\bmat}{\begin{pmatrix}}
\newcommand{\emat}{\end{pmatrix}}
\newcommand{\lb}{\left(}
\newcommand{\rb}{\right)}
\newcommand{\lsb}{\left[}
\newcommand{\rsb}{\right]}

\newcommand{\mb}{\mathbf} 
\newcommand{\mr}{\mathrm}

\usepackage[final]{hyperref} 
\hypersetup{
	colorlinks=true,       
	linkcolor=blue,        
	citecolor=blue,        
	filecolor=magenta,     
	urlcolor=blue         
}

\begin{document}
\title{Quantum description of Fermi arcs in Weyl semimetals in a magnetic field}

\author{Tim Bauer}
\affiliation{Institut f\"ur Theoretische Physik,
Heinrich-Heine-Universit\"at, D-40225  D\"usseldorf, Germany}
\author{Francesco Buccheri}
\affiliation{Dipartimento Scienza Applicata e Tecnologia, Politecnico di Torino, Corso Duca degli Abruzzi 24, 10129 Torino, Italy}
\affiliation{INFN Sezione di Torino, Via P. Giuria 1, 10125, Torino, Italy}
\author{Alessandro De Martino}
\affiliation{Department of Mathematics, City St~George's, University of London,
Northampton Square, EC1V OHB London, United Kingdom}
\author{Reinhold Egger}
\affiliation{Institut f\"ur Theoretische Physik,
Heinrich-Heine-Universit\"at, D-40225  D\"usseldorf, Germany}

\begin{abstract}
For a Weyl semimetal (WSM) in a magnetic field, a semiclassical description of the Fermi-arc surface state dynamics is usually employed for explaining various unconventional magnetotransport phenomena, e.g., Weyl orbits, the three-dimensional Quantum Hall Effect, and the high transmission through twisted WSM interfaces.
For a half-space geometry, we determine the low-energy quantum eigenstates for a four-band model of a WSM in a magnetic field perpendicular to the surface.  The eigenstates correspond to in- and out-going chiral Landau level (LL) states, propagating (anti-)parallel to the field direction near different Weyl nodes, 
which are coupled by evanescent surface-state contributions generated by all other LLs. 
These replace the Fermi arc in a magnetic field. 
Computing the phase shift accumulated between in- and out-going chiral LL states,
we compare our quantum-mechanical results to semiclassical predictions. 
We find quantitative agreement between both approaches. 
\end{abstract}
\maketitle

\section{Introduction}\label{sec1}

Two hallmark features of topological electronic systems are their anomalous magnetotransport properties and the existence of robust boundary states. In the rich material class of Weyl semimetals (WSMs) \cite{burkov2016topological,yan2017topological,armitage2018weyl,burkov2018weyl}, these distinct features manifest themselves in the chiral anomaly and in Fermi-arc surface states, respectively. WSMs are three-dimensional (3D) semimetals characterized by touching points of non-degenerate bands near the Fermi energy which are separated in the Brillouin zone.  These so-called Weyl nodes are effectively described by massless relativistic Weyl fermions with conserved chirality.    In the presence of electromagnetic fields, Weyl fermions exhibit the chiral anomaly which, on the level of the electronic band structure, implies the formation of a chiral zeroth Landau level (LL). This chiral LL state has a gapless linear dispersion along a direction determined by the chirality which is parallel or antiparallel to the magnetic field \cite{nielsen1983adler,hosur2013recent,burkov2015chiral,gorbar2018anomalous}.    On the other hand, Weyl nodes act as sources or sinks of Berry curvature and give rise to  non-trivial band topology \cite{volovik1987zeros,armitage2018weyl}.  Correspondingly, topological surface states emerge near the boundary of a WSM. Since the Nielsen-Ninomiya theorem requires Weyl nodes to come in pairs of opposite chirality \cite{nielsen1981absence}, the energy contour of these surface states must terminate at the projection of the bulk cones of two Weyl nodes on the surface Brillouin zone and form an open disjoint curve --- the Fermi arc.

\begin{figure}[t]
        \centering
        {\includegraphics[width=\columnwidth]{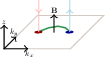}}
 \caption{Schematic sketch of the half-space WSM geometry (defined by $z\ge 0$) in a homogeneous magnetic field $\mb B=B \hat{\mb e}_z$ perpendicular to the surface.  The 3D WSM has two bulk Weyl nodes at momenta $\mb k=(\pm k_0,0,0)^T$, corresponding to the red and blue circles for their surface projections. For $B=0$, the surface projections are connected by a Fermi-arc surface state (green curve).  For $B\ne 0$, the low-energy bulk physics is dominated by the $n=0$ chiral LLs which have opposite chirality near different Weyl nodes
 (pink and light blue arrows show the respective propagation direction). 
 Fermions incoming from a bulk chiral LL enter the arc at the surface. After sliding along the arc
 to the opposite Weyl node surface projection (in a semiclassical picture), 
 they tunnel into the outgoing chiral LL \cite{potter2014quantum}.}
 \label{fig:1}
\end{figure}

Given the experimental observation of signatures for both the chiral anomaly and Fermi arcs \cite{xu2015discovery,zhang2016signatures,hasan2017discovery,ong2021experimental}, it is natural to ask how both phenomena conspire near the boundary of a semi-infinite WSM in a homogeneous magnetic field oriented perpendicular to the surface.  In a semiclassical picture, the presence of the Lorentz force implies that fermions slide along the Fermi arc connecting two Weyl node projections.    Due to the open nature of the Fermi-arc energy contour, no closed cyclotron orbit can form on the surface. Accordingly, fermions have to tunnel into the bulk upon reaching the chiral termination point of the Fermi arc \cite{potter2014quantum}, see Fig.~\ref{fig:1} for a schematic illustration in a half-space geometry. 
Since the only available bulk states at low energies are provided by the chiral LL states, semiclassics predicts that fermions will then move through the bulk and thereby escape from the surface. Consequently, Fermi-arc states acquire a finite lifetime in a perpendicular magnetic field $B$, and thus ultimately become unstable.  
Indeed, as we show in detail for the model studied below, for $B\ne 0$, no stable surface states exist anymore.  The true eigenstates for $B\ne 0$ have a component representing the zeroth-order chiral Landau levels in the bulk of the system. For $B\ne 0$, Fermi-arc electrons thus escape from the surface into the bulk via chiral Landau states, and therefore acquire a finite lifetime.  

Furthermore, in a WSM slab geometry (or in similar confined nanostructures), fermions in a chiral LL state move through the bulk and eventually tunnel into the opposite surface. There, they will traverse the corresponding opposite Fermi arc (in the semiclassical picture).    In the simplest scenario, the fermion subsequently occupies the chiral LL state with opposite chirality and travels back to the initial Fermi arc state. This closed trajectory resembles an exotic cyclotron orbit, commonly referred to as ``Weyl orbit''.
Such orbits are predicted to cause unconventional quantum oscillations in the magnetoconductivity, with a strong dependence on the sample thickness \cite{potter2014quantum,zhang2016quantum}. The initial prediction of Weyl orbits sparked excitement in the WSM community and led to a variety of subsequent theoretical proposals \cite{zhang2021cycling}, including non-local transport experiments \cite{parameswaran2014probing,baum2015current,hou2020nonlocal}, the detection of chiral separation \cite{gorbar2014quantum}, and the chiral magnetic effect \cite{zubkov2023weyl}.  Weyl orbits are also important ingredients for an unconventional so-called 3D Quantum Hall Effect (QHE) \cite{wang20173d,li20203d}. Both Shubnikov-de Haas oscillations due to Weyl orbits and the 3D QHE were thoroughly investigated in 
transport experiments for the Dirac semimetal Cd$_3$As$_2$ \cite{moll2016transport,zhang2017evolution,zheng2017recognition,zhang2019quantum,nishihaya2019quantized,nishihaya2021intrinsic,chen2021field},  
see also the review \cite{zhang2021cycling}.
   
In a Dirac semimetal (DSM), Weyl nodes of opposite chirality share the same position in momentum space but are stabilized by space group symmetries of the crystal \cite{armitage2018weyl}.
While the band structure is topologically trivial, pairs of Dirac cones can be connected by Fermi-arc surface states nevertheless.
 The semiclassical argument for the surface-bulk hybridization of Weyl orbits can be adapted to DSMs despite of the formally closed energy contour of surface states \cite{potter2014quantum}. While clear experimental evidence for the predicted signatures has been collected for Cd$_3$As$_2$  \cite{moll2016transport,zhang2017evolution,zheng2017recognition,zhang2019quantum,nishihaya2019quantized,nishihaya2021intrinsic}, their interpretation in terms of Weyl orbits remains debated \cite{zhang2021cycling}. In particular, thin films of Cd$_3$As$_2$ show an intricate dependence of the QHE on sample thickness \cite{nishihaya2019quantized,chang2021three}. Moreover, energy quantization due to Weyl orbits is difficult to distinguish from the trivial size quantization of confined bulk states \cite{nguyen2021quantum}. Similar arguments might also apply to the Weyl orbits reported in the non-centrosymmetric WSMs NbAs \cite{zhang2019ultrahigh}, TaAs \cite{nair2020signatures} and WTe$_2$ \cite{li2017evidence}. 
So far, no Weyl orbits have been observed in magnetic WSMs with broken time-reversal symmetry \cite{zhang2021cycling}.  However, recent experimental work on magnetic WSMs \cite{bernevig2022progress} and progress in quasiparticle interference experiments \cite{morali2019fermi} render near-future advances in Weyl-orbit physics for this class of materials likely.  These developments also motivated us to perform the present study.

A related exciting topic concerns twisted WSM interfaces \cite{dwivedi2018fermi,abdulla2021fermi,buccheri2022transport,kaushik2022transport} and tunnel junctions \cite{chaou2023magnetic}.  Upon twisting interfaces with respect to each other, theory predicts a 
Fermi-arc reconstruction, implying the existence of ``homo-chiral'' Fermi arcs connecting Weyl nodes of equal chirality \cite{abdulla2021fermi,buccheri2022transport}.
In a magnetic field perpendicular to the interface, incoming electrons in a chiral LL may traverse the homo-chiral Fermi arc and tunnel back into bulk states on the other side of the junction, thereby achieving (almost) perfect transmission.
Numerical transport simulations show good agreement with this semiclassical picture \cite{chaou2023magnetic,chaou2023quantum,breitkreiz2023fermi}.

Adopting the half-space geometry, we here study the fate  of Fermi-arc surface states in WSMs in a magnetic field. We construct a full quantum solution, going beyond  semiclassics. To this end, we study a four-band low-energy continuum model for a magnetic WSM. 
While we find analytical solutions of the eigenproblem for the DSM limit of two degenerate Weyl nodes, we develop a numerical approach (with a controlled cut-off procedure) to obtain the eigenstates for the WSM scenario. 
We find that the eigenstates with lowest energy are composed of in- and out-going chiral zeroth-order LLs which are coupled by evanescent states localized near the surface.  These are generated by all remaining higher-order LLs
and cause a phase shift between in- and outgoing chiral LLs. In a slab geometry, this phase shift is experimentally observable through magnetoconductivity oscillations \cite{potter2014quantum}. 
We compare our numerical results for the phase shift to semiclassical predictions by varying the energy $\varepsilon$ and a boundary parameter $\alpha$ encoding the arc curvature in the surface momentum plane.  In addition, the energy derivative of the phase shift determines the Fermi-arc lifetime which is finite for 
$B\ne 0$.  We show how the lifetime depends on key parameters such as $\alpha$, $\varepsilon,$ and $B$,
and compare it to the semiclassical traversal time across the Fermi arc.  

The remainder of this paper is structured as follows. In Sec.~\ref{sec2} we discuss the continuum WSM model employed here, derive boundary conditions for the half-space geometry, and present the surface state spectrum at zero magnetic field.
We include the magnetic field in Sec.~\ref{Landau} and construct the eigenstates in the half-space geometry.
In addition, we consider the limit of a DSM and obtain analytical solutions in several limiting cases.
Subsequently, we derive the corresponding semiclassical predictions in Sec.~\ref{semiclassics} and compare them with our 
quantum-mechanical results for the phase shift and for the Fermi-arc lifetime.  The paper closes with concluding remarks in Sec.~\ref{discussion}.  Details of our calculations can be found in several appendices.  
A derivation of $B=0$ Fermi-arc surface states is given in App.~\ref{surface_states}.  
Their spin and current structure is discussed in App.~\ref{spinArc}.
We validate our numerical approach for finite magnetic fields in App.~\ref{appnumerics}, and discuss the Goos-H\"anchen
effect for the present setup in App.~\ref{GHshift}.
Throughout this paper, we use units with Fermi velocity $v_F=1$ and put $\hbar=1$.

\section{WSM in half space}\label{sec2}

In this section, we discuss the four-band WSM model employed in our study.  The 3D model (in the absence of a magnetic field) is 
introduced in Sec.~\ref{sec2a}.  We then discuss the half-space geometry in terms of boundary conditions in Sec.~\ref{sec2b}.
The Fermi-arc surface states for $B=0$ are specified in Sec.~\ref{sec2c}, see also App.~\ref{surface_states} and App.~\ref{spinArc} for further details.

\subsection{Model}\label{sec2a}

We study a four-band continuum WSM model which in 3D space, with conserved momentum $\mb k=(k_x,k_y,k_z)^T$, is defined by the Hamiltonian 
\cite{armitage2018weyl}
\be \label{eq:model}
        H(\mb k)=\mb k\cdot\boldsymbol{\sigma}\tau^z+k_0\sigma^x\tau^0,
\ee
where $k_0\geq 0$ is the only free parameter. This parameter determines the distance between the Weyl nodes in momentum space. In Eq.~\eqref{eq:model},
$\sigma^\mu$ and $\tau^\mu$ are Pauli matrices acting on effective spin and orbital degrees of freedoms, respectively, where $\mu=0$ refers to the identity and $\mu=x,y,z$ otherwise. We use $\boldsymbol{\sigma}=(\sigma^x,\sigma^y,\sigma^z)$.
While the limit $k_0=0$ describes a degenerate Dirac cone centered at $\mb k= \mb 0$, i.e., a Dirac semimetal, the model exhibits two separated Weyl nodes at momenta $\mb k=\pm k_0\hat{\mb e}_x$ for $k_0>0$. Due to the block-diagonal structure of $H$, these Weyl nodes are decoupled. Their conserved chirality $\chi$ is associated with the orbital degree of freedom, namely the eigenvalues $\chi=\pm 1$ of $\tau^z$.
    
We note that adding a mass term $H'=m\sigma^0\tau^x$ in Eq.~\eqref{eq:model} couples the Weyl nodes.   However, for $m<k_0$, the Weyl nodes are thereby only shifted in momentum space and the low-energy description is not changed in an essential manner \cite{armitage2018weyl}. We thus put $m=0$ throughout this work. The
two Weyl nodes are then fully decoupled \emph{in the bulk}. This key simplification allows us to obtain explicit 
results in a finite magnetic field. Importantly, in our approach, the boundary conditions will couple both Weyl nodes. 
Furthermore, while Eq.~\eqref{eq:model} formally describes a type-I WSM with broken time-reversal symmetry and the minimum number of two Weyl nodes, we expect our arguments to apply to any WSM involving a pair of two isolated type-I Weyl nodes.

Below, we use the standard representation of Pauli matrices. States are written in the eigenbasis of  $\sigma^z$ and $\tau^z$, i.e., $\ket{\chi}_{\boldsymbol{\tau}}$ for chirality $\chi=\pm 1$ and $\ket{\sigma}_{\boldsymbol{\sigma}}$ for spin $\sigma\in\{\uparrow,\downarrow\}$, with the spinor representations
\bea\nonumber
\ket{+}_{\boldsymbol{\tau}}&=\bmat 1\\0\emat_{\boldsymbol{\tau}},\quad 
\ket{-}_{\boldsymbol{\tau}}&=\bmat 0\\1\emat_{\boldsymbol{\tau}},\\
\ket{\uparrow}_{\boldsymbol{\sigma}}&=\bmat 1\\0\emat_{\boldsymbol{\sigma}},\quad 
\ket{\downarrow}_{\boldsymbol{\sigma}}&=\bmat 0\\1\emat_{\boldsymbol{\sigma}}.
\eea
    
\subsection{Half-space geometry and boundary conditions}\label{sec2b}

We next proceed to the half-space geometry defined by $z\geq 0$, where we have a planar boundary at $z=0$ as illustrated in Fig.~\ref{fig:1}.    Since the surface projections of the two Weyl nodes  are separated, topological Fermi-arc surface 
states with an open energy contour connecting the projected nodes arise \cite{armitage2018weyl}. 
Before solving for the Fermi arcs, we first derive the general boundary condition from the constraint of Hermiticity of 
the Hamiltonian.  For relativistic continuum models, such an approach typically allows for a few free parameters with physical implications on the surface state dispersion \cite{witten2016three,hashimoto2017boundary,thiang2021spectral,Buccheri2022a}.
We note that a more realistic modeling of the boundary might include band bending near the surface which can drastically change the dispersion \cite{Buccheri2024,li2015spiraling}.
    
For a derivation that is also valid in the presence of a finite magnetic field, we switch to    real space  by using the substitution  $\mb k\rightarrow-i\nabla_{\mb r}$ in Eq.~\eqref{eq:model}. Following standard arguments \cite{witten2016three,hashimoto2017boundary,thiang2021spectral}, we impose  $\braket{\Psi_1|H\Psi_2}-\braket{H\Psi_1|\Psi_2}=0$ for arbitrary states $\Psi_1$ and $\Psi_2$ in the half-space geometry to infer a  \emph{sufficient} boundary condition,
\be\label{cond1}
 \Psi_1^\dagger(\mb{r}_{\perp},z=0)j^z\Psi_2(\mb{r}_{\perp},z=0)=0.
\ee
 Here, $\mb{r}_\perp=(x,y)^T$ is the in-plane position and
 $j^z=\sigma^z\tau^z$ is the $z$-component of the relativistic fermion particle current operator, $\mb j=\boldsymbol{\sigma}\tau^z$.
Physically, Eq.~\eqref{cond1} thus prohibits any local current flowing through the surface.  
This condition is ensured for states that satisfy boundary conditions of the form
\be \label{eq:BCM}
  M\Psi(\mb{r}_{\perp},z=0)=\Psi(\mb{r}_{\perp},z=0),
\ee
where $M$ is an operator with the properties
\be
j^zM=-M^\dagger j^z,\quad M^2=\mathbbm{1},
\ee
where $\mathbbm{1}=\sigma^0\tau^0$ is the identity. In Eq.~\eqref{eq:BCM},  we assumed a local boundary condition where the matrix $M$ does not depend on the in-plane position $\mb{r}_\perp$.  
To parameterize all possible choices of $M$, we define the operators
\be \label{eq:Mtausigma}
 M^{\boldsymbol{\tau}}_\gamma=\tau^x\cos{\gamma}+\tau^y\sin{\gamma},\quad M^{\boldsymbol{\sigma}}_\delta=\sigma^x\cos{\delta}+\sigma^y\sin{\delta},
\ee
in orbital and spin space, respectively. The most general Hermitian parameterization
then involves four real parameters ($\alpha,\beta,\gamma,\delta)$ \cite{Faraei2018},
\bea \label{eq:M_general}
M_{\alpha\beta\gamma\delta}&=&\cos{\alpha}\lb \sigma^zM^{\boldsymbol{\tau}}_\gamma\cos{\beta}
 +\sigma^0M^{\boldsymbol{\tau}}_{\gamma-\pi/2}\sin{\beta}\rb\nonumber\\
 &+&\sin{\alpha}\lb\tau^0 M^{\boldsymbol{\sigma}}_\delta\cos{\beta}+ \tau^z M^{\boldsymbol{\sigma}}_{\delta-\pi/2} \sin{\beta}\rb.
\eea
An equivalent parameterization was found in Refs.~\cite{akhmerov2007detection,akhmerov2008boundary} in the context of  graphene monolayers.   On general grounds, the number of free parameters in the boundary condition increases with the number of higher-energy bands in the model Hamiltonian \cite{bovenzi2018twisted,burrello2019field}.  Consequently, the low-energy spectrum is not expected to change significantly when varying parameters within certain submanifolds of the full parameter space.  Below, we do not exploit the complete parameter freedom in Eq.~\eqref{eq:M_general} but instead focus on a simple one-parameter boundary matrix $M$ allowing us to describe curved Fermi arcs.

\subsection{Boundary spectrum at zero magnetic field}\label{sec2c}

\begin{figure}[t]
\centering
\includegraphics[width=\columnwidth]{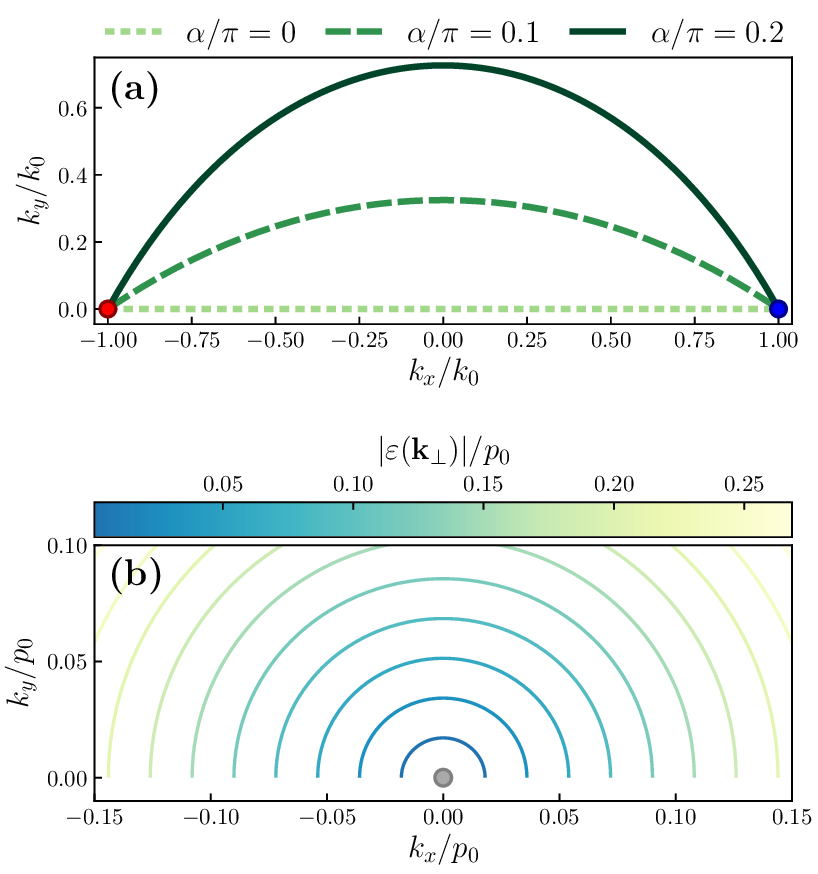}
\caption{Surface-state spectrum of the four-band model \eqref{eq:model} in a half space for $B=0$. 
(a) Zero-energy Fermi-arc contours in the $k_x$-$k_y$ plane as described by Eq.~\eqref{eq:contour} for different boundary
 parameters $\alpha$, see also Fig.~\ref{fig:1}. (b) Surface-state contour-plot in the $k_x$-$k_y$ plane for a DSM ($k_0\to 0$) with $\alpha/\pi=0.1$ for different energies $\varepsilon(\mb k_\perp)$ 
 as indicated by the color bar, using a fixed scale $p_0$ as reference. 
 The surface-state termination points result from the condition in Eq.~\eqref{eq:constraint_Dirac}.}
\label{fig:2}
\end{figure}
    
Given the boundary condition \eqref{eq:BCM} with the general parameterization \eqref{eq:M_general}, we next 
construct physical Fermi arc solutions for $B=0$, which are labeled by the conserved in-plane momentum $\mb k_\perp=(k_x,k_y)^T$.  The simplest approach is to choose a block diagonal matrix $M$, e.g., the parametrization $M_{\frac{\pi}{2},0,0,\delta}$ in Eq.~\eqref{eq:M_general}, which allows one
to solve the problem for both Weyl nodes separately.
The Fermi arc of a single Weyl node, which after a shift of $k_x$ is effectively described by $H_\chi=\chi\mb{k}\cdot\boldsymbol{\sigma}$, then becomes a semi-infinite line which terminates at the Weyl cone projection at an angle determined by the parameter $\delta$ \cite{witten2016three,hashimoto2017boundary,thiang2021spectral}.
The resulting surface-state spectrum of the four-band WSM model thus yields two semi-infinite arcs, 
in contrast to physical Fermi arcs which are open curves connecting both Weyl cone projections.
Here, we will use boundary conditions which couple different Weyl nodes. This approach is especially
convenient for $B\ne 0$.

To this end, we consider off-diagonal boundary matrices $M$ satisfying $[M,\sigma^0\tau^z]\neq 0$. The resulting boundary conditions couple the Weyl nodes \emph{at the surface} \cite{Faraei2018}. This picture is analogous to the one in Ref.~\cite{akhmerov2008boundary},  where armchair edges in graphene monolayers are modeled by boundary conditions that are non-diagonal in the valley degree of freedom.   
Below, we assume the boundary condition \eqref{eq:BCM} with
\be  \label{eq:Malpha}
        M_\alpha=M_{\alpha,0,0,0}=\sigma^z\tau^x\cos{\alpha}+ \sigma^x\tau^0\sin{\alpha}.
\ee  
The parametrization \eqref{eq:Malpha} is a simple choice that allows us to construct physical Fermi arcs with a curvature in the surface momentum plane
controlled by the single parameter $\alpha$.  A curved arc is then symmetric 
under mid-point reflection, and a straight arc is found for $\alpha=0$ (mod $\pi$).  
As discussed in App.~\ref{surface_states}, the particular choice of $M_\alpha$ in Eq.~\eqref{eq:Malpha} is motivated by the observation that a straight arc requires $[M,j^y]=0$, where $j^y=\sigma^y\tau^z$ is the in-plane current operator along $\hat{\mb e}_y$.
For a microscopic analysis of a specific material, one may instead employ boundary matrices $M$ containing more parameters (while still coupling both Weyl nodes), 
possibly guided by numerical calculations for lattice models.   For simplicity, however, we focus on the one-parameter family of matrices in Eq.~\eqref{eq:Malpha}. 
In App.~\ref{surface_states}, we derive the corresponding $B=0$ surface-state spectrum presented next. 

We find that a physical Fermi-arc contour at energy $\varepsilon$ is given by $k_y=q_\alpha(\varepsilon,k_x)$, where 
\be\label{eq:contour}
  q_\alpha(\varepsilon,k_x)=\frac{(\varepsilon\sin{\alpha}-k_0)(\varepsilon-k_0\sin{\alpha})-k_x^2\sin{\alpha}}{\cos{\alpha}\sqrt{(\varepsilon\sin{\alpha}-k_0)^2-k_x^2\sin^2{\alpha}}}.
\ee
At zero energy, the contour terminates at both Weyl node surface projections ($k_x=\pm k_0$, $k_y=0$). 
The termination points $k_x=\pm k_{\varepsilon\alpha}$ for $\varepsilon\ne 0$ are implicitly given by
\be\label{eq:norm_constraint_main}
    \varepsilon^2={\lb|k_x|-k_0\rb^2+[q_\alpha(\varepsilon,k_x)]^2}.
\ee 
Assuming $|\varepsilon|\ll k_0$ and expanding Eq.~\eqref{eq:contour} to lowest order in $\varepsilon/k_0$, we estimate
\be\label{eq:termination_point}
    k_{\varepsilon\alpha}=k_0-\frac{4\varepsilon\sin{\alpha}}{3-\cos{(2\alpha)}}.
\ee
Moreover, we find the low-energy dispersion relation
\begin{align}\label{eq:low_dispersion}\nonumber
\varepsilon(\mb k_\perp)&\simeq \frac{k_0^2-k_x^2\sin^2{\alpha}}{k_0^2-k_x^2\sin^4{\alpha}}\bigg(\frac{k_0^2-k_x^2}{k_0}\sin{\alpha}\\&-\sqrt{k_0^2-k_x^2\sin^2{\alpha}} \, \frac{k_y}{k_0}  \cos{\alpha}\bigg).
\end{align}
For $\alpha=0$, the above expressions are exact and describe a straight arc with  $\varepsilon(\mb k_\perp)=-k_y$.
For $\alpha\ne 0$, the arc is curved in the surface momentum plane  as illustrated in Fig.~\ref{fig:2}(a). 
We note that the Fermi arc for $\alpha\to\pi-\alpha$ with the same energy $\varepsilon$ follows
by reflection with respect to the $k_x$-axis. In effect, this transformation yields the Fermi arc for
the same boundary condition but in the opposite half-space $z\le 0$, see App.~\ref{surface_states}.
We briefly discuss the in-plane spin and current densities associated with Fermi-arc states in App.~\ref{spinArc}.
    
Next, we turn to the limit $k_0\to 0$ describing the low-energy theory of a DSM with a single degenerate cone. While the band structure is now topogically trivial, surface states may nonetheless exist.  Such states become important for $B\ne 0$, see Sec.~\ref{semiclassics}.  Deferring technical details to App.~\ref{surface_states}, we find topologically trivial surface states with the dispersion relation
\be        \label{eq:dispersion_Dirac}
\varepsilon_\pm(\mb k_\perp)=\pm\sqrt{k_x^2+k^2_y\cos^2{\alpha}},
\ee
which only exist if the condition 
\be \label{eq:constraint_Dirac}    
k_y\sin{\alpha}>0
\ee
is satisfied. 
In particular, there neither are surface state solutions for $\alpha=0$ (mod $\pi$), corresponding to a straight arc for finite $k_0$, 
nor for $\varepsilon=0$.  Below, we focus on those two cases for analytical results.
For finite $\alpha$ or $\varepsilon$, however, surface states emerge which form open energy contours. These contours shrink with decreasing energy, see Fig.~\ref{fig:2}(b).  

\section{Half space in a magnetic field}\label{Landau}

In this section, we include the magnetic field $\mb B=B\hat{\mb e}_z$ with $B>0$ and study the WSM model in Sec.~\ref{sec2} for the half-space
geometry using the boundary condition \eqref{eq:BCM} defined by the matrix $M_\alpha$ in Eq.~\eqref{eq:Malpha}.  The parameter  $\alpha$ determines the curvature of the $B=0$ Fermi arc solutions.  
In Sec.~\ref{sec3a}, we briefly review the eigenstates for the infinite 3D problem. In Sec.~\ref{sec3b}, we then turn to the half-space problem and construct the low-energy quantum-mechanical eigenstates.

\subsection{Landau quantization}\label{sec3a}

We start with the free-space WSM model and we incorporate the homogeneous magnetic field $\mb{B}$ by minimal coupling, $\mb k\to -i\nabla_{\mb r} +\frac{e}{c}\mb{A}$,  where $e>0$ is the (absolute value of the) electron charge and
$c$ is the speed of light.  For convenience, we choose the Landau gauge, $\mb{A}=-By\hat{\mb e}_x$, where Eq.~\eqref{eq:model} gives
\be   \label{eq:Hmagnetic}
H=-i\nabla_{\mb r}\cdot\boldsymbol{\sigma}\tau^z-\frac{y}{\ell_B^2}\sigma^x\tau^z+k_0\sigma^x\tau^0  =\bmat H_+&0\\0&H_-\emat_{\boldsymbol{\tau}}
\ee
with the magnetic length $\ell_B=\sqrt{c/eB}$.   Note that the chosen gauge retains translation invariance along $\hat{\mb e}_x$. The momentum component $k_x$ therefore remains a good quantum number.
Below, we consider the orbital magnetic field only and neglect the Zeeman effect by following standard arguments \cite{nielsen1983adler,hosur2013recent,burkov2015chiral,gorbar2018anomalous}. Including a Zeeman term, say, of the form $H_Z\propto \sigma^z\tau^0$ shifts the position of the Weyl nodes in the $xz$ plane. While such an effect can be readily taken into account, we assume here for simplicity that its contribution is insignificant compared to the second term in Eq.~\eqref{eq:model}.

When solving for LL solutions, it is convenient to consider the Weyl nodes separately. A single Weyl node with chirality $\chi=\pm 1$ and momentum $\mb k=-\chi k_0\hat{\mb e}_x$ is described by the Hamiltonian $H_\chi$ in spin space using the block diagonal form in Eq.~\eqref{eq:Hmagnetic}.  For given $\chi$ and $k_x$, we define the bosonic ladder operator
\be
a^\dagger_\chi=\frac{\ell_B}{\sqrt{2}}\lb \frac{y}{\ell_B^2}-k_x-\chi k_0-\partial_y\rb,
\ee
with the commutator $[a_\chi,a_\chi^\dagger]=1$.
The transverse part of $H_\chi=-i\chi\partial_z\sigma^z+H^\perp_\chi$ is thereby written as
\be
        H_\chi^\perp=-\frac{\chi}{\sqrt{2}\ell_B}\lsb\lb a_\chi+a_\chi^\dagger\rb\sigma^x+i\lb a_\chi-a_\chi^\dagger\rb\sigma^y\rsb.
\ee
In the infinite 3D system (without boundary), the momentum component $k_z$ is also conserved. It is then straightforward to obtain the well-known relativistic LLs $\varepsilon^\chi_{nk_z}$ labeled by non-negative integer $n\in\mathbbm{N}_0$ \cite{armitage2018weyl},
\be\label{eq:bulk_LL}
    \varepsilon^\chi_{0,k_z}=-\chi k_z,\quad \varepsilon^\chi_{\pm,n>0,k_z}=\pm\chi\sqrt{\frac{2n}{\ell_B^2}+k_z^2}.
\ee
Here, $\varepsilon^\chi_{0,k_z}$ is the dispersion of the gapless chiral LL, while $n> 0$ correspond to higher-order gapped LL states.    Eigenstates are expressed in terms of harmonic oscillator eigenfunctions,
\be\label{harmonicmodes}
        \varphi_n(y) =\frac{ H_n(y/\ell_B) }{ \sqrt{2^n n! \sqrt{\pi}\, \ell_B}}\,{\mr e}^{-\frac{1}{2}{(y/\ell_B)^2}},
\ee
where $H_n$ is the $n$th-order Hermite polynomial. 
Writing $a_\chi^\dagger a_\chi\varphi_n^\chi=n\varphi_n^\chi$,
the wave functions 
\be\label{varphin}
\varphi_n^\chi(y)=\varphi_n(y-\ell_B^2k_x-\chi \ell_B^2 k_0)
\ee
incorporate a shift with respect to the Weyl node position.
In anticipation of the half-space geometry, we 
label the solutions $\ket{\psi^\chi_{n\varepsilon}}$ of $H_\chi\ket{\psi^\chi_{n\varepsilon}}=\varepsilon\ket{\psi^\chi_{n\varepsilon}}$ in terms of energy $\varepsilon$ instead of $k_z$. 
The chiral LL with $n=0$ is then described by 
\be \label{eq:LLchiral}
 \psi^\chi_{0,\varepsilon}(y,z)=\frac{\mr{e}^{-i\chi\varepsilon z}}{\sqrt{2\pi}}\bmat 0\\
 \varphi_{0}^\chi(y)\emat_{\boldsymbol{\sigma}}.
\ee
Since $k_x$ is conserved, we keep plane-wave factors $e^{ik_x x}$ and the $k_x$-dependence of observables 
implicit below.
Similar expressions as Eq.~\eqref{eq:LLchiral} hold for the wave functions of $n> 0$ bulk LLs \cite{nielsen1983adler}.

In the following, we focus on the \emph{ultra-quantum regime}, $|\varepsilon|<{\sqrt{2}}/{\ell_B}$.
While $n> 0$ bulk LLs do not exist in this regime, it is possible to construct \emph{evanescent} solutions in the half-space geometry by solving the eigenproblem for imaginary momentum $k_z=i\kappa$ with $\kappa=\kappa_{n\varepsilon}>0$.    The evanescent solution for $n>0$ is given by
\be \label{eq:LLevanes}
\psi^\chi_{n\varepsilon}(y,z)=\sqrt{\kappa_{n\varepsilon}}\,{\mr{e}^{-\kappa_{n\varepsilon} z}}\bmat \chi\mr{e}^{i\chi\gamma_{n\varepsilon}}\varphi_{n-1}^\chi(y)\\\varphi_{n}^\chi(y)\emat_{\boldsymbol{\sigma}},
\ee
with the inverse penetration length
\be
\kappa_{n\varepsilon}=\sqrt{2n/\ell_B^2-\varepsilon^2}
\ee
and the phase $\gamma_{n\varepsilon}$ defined by
\be
\mr{e}^{i\gamma_{n\varepsilon}}=-\frac{\ell_B}{\sqrt{2n}}(\varepsilon+i\kappa_{n\varepsilon}).
\ee
One can rationalize the appearance of this complex phase factor by noticing that evanescent solutions do not carry any current along $\hat{\mb e}_z$, i.e., $\braket{\psi^\chi_{n\varepsilon}|\sigma^z|\psi^\chi_{n\varepsilon}}=0$ for $n>0$.
    
\subsection{Half-space geometry}\label{sec3b}

\subsubsection{Coupling of Weyl nodes at the boundary}

We now proceed to the half-space geometry $z\ge 0$ sketched in Fig.~\ref{fig:1}, see Sec.~\ref{sec2} for the $B=0$ case. We first rewrite the boundary condition \eqref{eq:BCM} with the matrix $M_\alpha$ in Eq.~\eqref{eq:Malpha} as
\be        \label{eq:BCV}
\mathbbm{V}_\alpha(z)\Psi(\mb r)=0,\quad \mathbbm{V}_\alpha(z)=\delta(z)\lb\mathbbm{1}-M_\alpha\rb.
\ee
Our \emph{Ansatz} for solving Eq.~\eqref{eq:BCV} is a superposition of all eigenstates of $H$ in Eq.~\eqref{eq:Hmagnetic} with given  $\varepsilon$ and 
 $k_x$.  We focus on the ultra-quantum regime $|\varepsilon|<\sqrt{2}/\ell_B$, where 
$n> 0$ LL states only contribute through evanescent-state solutions in Eq.~\eqref{eq:LLevanes}.  
Combining the results of Eqs.~\eqref{eq:LLchiral} and \eqref{eq:LLevanes} gives  
\be \label{eq:superposition}
\ket{\Psi_\varepsilon}=\sum_{\chi=\pm}\sum_{n\geq 0} 
c^\chi_{n\varepsilon}\ket{\psi^\chi_{n\varepsilon}}_{\boldsymbol{\sigma}}\ket{\chi}_{\boldsymbol{\tau}},
\ee
where the $c^\chi_{n\varepsilon}$ are complex coefficients which have to be determined.
Equation \eqref{eq:BCV} states that $\ket{\Psi_\varepsilon}$ is element of the kernel of  $\mathbbm{V}_\alpha$.    Matrix elements of this operator, restricted to the 
subspace with fixed energy $\varepsilon$, are of the form
\be  \label{eq:matrixelements}
\lsb V_\alpha(\varepsilon)\rsb^{\chi,\chi'}_{n,n'}={}^{}_{\boldsymbol{\tau}}\bra{\chi}
\, {}^{}_{\boldsymbol{\sigma}}\bra{\psi_{n\varepsilon}^\chi} \mathbbm{V}_\alpha(z)
\ket{\psi^{\chi'}_{n'\varepsilon}}_{\boldsymbol{\sigma}}\ket{\chi'}_{\boldsymbol{\tau}}.
\ee
For convenience, we rescale them as
\be \label{eq:rescaled_V}
 [\hat{V}_\alpha(\varepsilon)]^{\chi,\chi'}_{n,n'}=\lb \kappa_{n\varepsilon}\kappa_{n'\varepsilon}\rb^{-\frac{1}{2}}\lsb V_\alpha(\varepsilon)\rsb^{\chi,\chi'}_{n,n'}
\ee 
with $\kappa_{0,\varepsilon}=1/2\pi$.
Matrix elements between a chiral $n=0$ LL and $n\ge 0$ LLs with equal chirality $\chi$ are given by
\be
        [\hat{V}_\alpha(\varepsilon)]^{\chi,\chi}_{0,n}=\delta_{n,0}-\chi\mr{e}^{i\chi\gamma_{n\varepsilon}}\delta_{n,1}\sin{\alpha},
\ee
while for $n,n'>0$, we find
\begin{align}\nonumber
        [\hat{V}_\alpha(\varepsilon)]^{\chi,\chi}_{n,n'}&=2\delta_{n,n'}-\chi\mr{e}^{i\chi\gamma_{n+1,\varepsilon}}
        \delta_{n,n'-1}\sin{\alpha}\\
        &-\chi\mr{e}^{-i\chi\gamma_{n\varepsilon}}\delta_{n,n'+1}\sin{\alpha}.
\end{align}
Matrix elements for opposite chiralities resemble the coupling of Weyl nodes in terms of the boundary condition.
 For $n\geq 0$, we obtain
\be        \label{eq:matrixpm1}
     [\hat{V}_\alpha(\varepsilon)]^{\chi,-\chi}_{0,n}=\braket{\varphi^\chi_0|\varphi^{-\chi}_n}\cos{\alpha}.
\ee
Finally, for $n,n'>0$, we find
\begin{align}\label{eq:matrixpm2}
        [\hat{V}_\alpha(\varepsilon)]^{\chi,-\chi}_{nn'}&=\cos{\alpha}\lb\mr{e}^{-i\chi(\gamma_{n\varepsilon}+\gamma_{n'\varepsilon})}\braket{\varphi^\chi_{n-1}|\varphi^{-\chi}_{n'-1}}\right.\\ \nonumber
        &\left.+\braket{\varphi^\chi_{n}|\varphi^{-\chi}_{n'}}\rb.
\end{align}
The overlap $\braket{\varphi^{\chi}_n|\varphi^{-\chi}_m}$ involves shifted harmonic oscillator eigenfunctions
associated with different Weyl nodes.  Performing the integration for $n\geq m$ yields \cite{gradshteyn2014table}
\begin{align}\label{eq:overlaps}
        \braket{\varphi^{-}_n|\varphi^{+}_m}&=\int dy\,\varphi_n(y-l_B^2k_0)\varphi_m(y+l_B^2k_0)\\ \nonumber
        &=\sqrt{2^{n-m}\frac{m!}{n!}}\lambda_B^{n-m}L_m^{n-m}(2\lambda_B^2)\mr{e}^{-\lambda_B^2},
\end{align}
with the dimensionless quantity
\be\label{lambdaB}
\lambda_B=k_0\ell_B,
\ee
which measures the decoupling of the Weyl nodes by the magnetic field. 
In Eq.~\eqref{eq:overlaps}, $L^m_n$ is a generalized $n$th-order Laguerre polynomial.
In App.~\ref{appnumerics}, we describe a recursion relation allowing for the numerically efficient computation of 
the overlaps in Eq.~\eqref{eq:overlaps}.    The remaining terms follow from the relation
 $\braket{\varphi^{+}_n|\varphi^{-}_m}=(-1)^{n-m}\braket{\varphi^{-}_n|\varphi^{+}_m}$.
We note that the overlaps allow for a perturbative treatment in the large-field limit $\lambda_B\ll 1$.
Let us also mention in passing that similar expressions appear when computing matrix elements of the bulk
mass term $m\sigma^0\tau^x$. In that case, the coupling opens a gap in the dispersion of the chiral LLs of the order of  
$\braket{\varphi_0^{+}|\varphi_0^{-}}=\mr{e}^{-\lambda_B^2}$.
This result is consistent with the WKB approximation for a two-band WSM model with two Weyl nodes \cite{saykin2018landau,chan2017emergence}. 

In any case, convergence of the overlaps $\lim_{n\to\infty}\braket{\varphi^{+}_n|\varphi^{-}_m}=0$ is ensured for arbitrary $\lambda_B$.  This fact justifies the introduction of a cut-off $N$ for the LL index, $n<N$, 
reducing the numerical solution of the  boundary problem to a linear algebra problem,
\be        \label{eq:linearsystem}
        V_\alpha(\varepsilon){\mb c}_\varepsilon=\mb 0,
\ee
where $V_\alpha(\varepsilon)$ is a $2N\times 2N$ matrix formed by the matrix elements \eqref{eq:matrixelements} of the lowest $N$ LLs and $\mb c_\varepsilon$ is a vector containing the corresponding coefficients $c^\chi_{n\varepsilon}$. 
The numerical solution of Eq.~\eqref{eq:linearsystem} then determines the eigenstates of the WSM in the half-space geometry for  $B\ne 0$.    In App.~\ref{appnumerics}, we carefully verify the controlled nature of the above cut-off procedure and the accuracy of the boundary condition. 

Due to current conservation, coefficients with the same $(n,\varepsilon,k_x)$ 
but different chiralities have the same absolute value,
$|c_{n\varepsilon}^{+}|=|c_{n\varepsilon}^{-}|$. 
In particular, we are interested in the phase shift $\theta_{\alpha}(\varepsilon)$ between in- and outgoing chiral $n=0$ Landau states,  
\be\label{eq:phaseshift}
{c_{0,\varepsilon}^{-}}=\mr{e}^{i\theta_\alpha(\varepsilon)}{c_{0,\varepsilon}^{+}}.
\ee
We note that all phases below are defined only modulo $2\pi$.
The phase shift $\theta_\alpha(\varepsilon)$ depends on the global phase choices for the basis states in Eq.~\eqref{eq:superposition}. 
While for a fixed phase choice, $\theta_\alpha(\varepsilon)$ is formally gauge invariant, observable quantities must also be independent of the phase choice. 
Full gauge invariance is ensured below by only considering phase shift differences, 
$\Phi=\theta_{\alpha'}(\varepsilon')-\theta_{\alpha}(\varepsilon)$.  
When combined with the corresponding phase shift on the opposite surface in a slab geometry, one can infer 
the magnetoconductivity oscillation period of the corresponding Weyl orbit from Eq.~\eqref{eq:phaseshift} \cite{potter2014quantum}. We compare our quantum-mechanical results for $\Phi$ to 
semiclassical estimates in Sec.~\ref{semiclassics}.

We note that for a straight arc at zero energy, $\alpha=0$ (mod 2$\pi$) and $\varepsilon=0$, with the basis choice in Eq.~\eqref{eq:superposition}, one finds
\be \label{pilock}
\theta_{\alpha=0}(\varepsilon=0)= \pi.
\ee 
We verify Eq.~\eqref{pilock} by evaluating the boundary condition at $y=\ell_B^2k_x$, where $\varphi^+_n(\ell_B^2k_x)=(-1)^n \varphi^-_n(\ell_B^2k_x)$. 
By virtue of $|c^+_{n\varepsilon}|=|c^-_{n\varepsilon}|$ and the boundary condition, we then arrive at $c_{n,0}^+=(-1)^{n+1}c_{n,0}^-$, and thus at Eq.~\eqref{pilock}.

Since eigenstates in the half-space geometry can be written in the form \eqref{eq:superposition}, a nontrivial $y$-dependence arises since the separation between Weyl nodes in momentum space appears in the argument of Eq.~\eqref{varphin}. As shown in App.~\ref{GHshift}, this observation implies that an electron incident on the surface undergoes a shift (assuming $\varepsilon>0$)
\begin{equation}\label{GHdeltay}
  \delta y=-2\ell_{B}^{2}k_{0}  
\end{equation}
in the $y$-direction. This effect can be interpreted semiclassically in terms of chiral transport associated to Fermi arcs, see App.~\ref{spinArc}. 
    
\subsubsection{Dirac semimetal}\label{dsm}
   
\begin{figure}[t]
\centering
\includegraphics[width=\columnwidth]{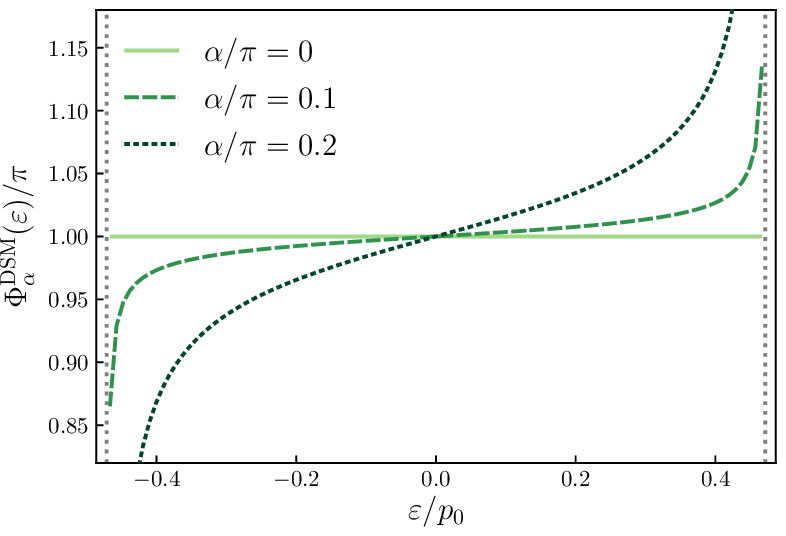}
 \caption{Gauge invariant phase shift $\Phi^{\rm DSM}_\alpha(\varepsilon)$ vs energy $\varepsilon$ for chiral LLs in a DSM ($k_0\to 0$) for several values of $\alpha$, 
 see Eq.~\eqref{dsmphaseshift}.  The shown results were obtained by a numerical solution of the quantum-mechanical problem. 
 We use $p_0\ell_B=3$ with a reference scale $p_0$, where $\varepsilon$ is shown in units of $p_0$ (with $v_F=1$).  The ultra-quantum regime $|\varepsilon|<\sqrt2/\ell_B$ is indicated by  vertical dotted lines.
 }
 \label{fig:3}
\end{figure}
    
In order to identify contributions  to the phase shift \eqref{eq:phaseshift} picked up by fermions traversing the Fermi arc in Sec.~\ref{semiclassics}, let us briefly consider the analogous problem in the DSM limit $k_0\to 0$.    The corresponding linear system follows from Eq.~\eqref{eq:linearsystem} by inserting diagonal overlaps $\braket{\varphi^{-}_n|\varphi^{+}_m}=\delta_{nm}$ in Eqs.~\eqref{eq:matrixpm1} and \eqref{eq:matrixpm2}.  For analytical results, we focus on cases without topologically trivial surface states for $B=0$, i.e., we consider either $\alpha=0$ (mod $\pi$) or $\varepsilon=0$. 

First, for $\alpha=0$ (mod~$2\pi$), it is straightforward to show that the boundary condition \eqref{eq:BCM} with $M_0=\sigma^z\tau^x$ is satisfied by antisymmetric superpositions of chiral LLs,
\be        \label{eq:Dirac_straight}
        \ket{\Psi_\varepsilon}=\frac{1}{\sqrt{2}}\lb\ket{\psi^{+}_{0,\varepsilon}}_{\boldsymbol{\sigma}}\ket{+}_{\boldsymbol{\tau}}-\ket{\psi^{-}_{0,\varepsilon}}_{\boldsymbol{\sigma}}\ket{-}_{\boldsymbol{\tau}}\rb.
\ee
With the above basis choice, we then obtain the phase shift $\theta_{\alpha=0}(\varepsilon)=\theta^{\rm DSM}_{\alpha=0}(\varepsilon)= \pi$ for arbitrary $\varepsilon$.  Similarly, one finds $\theta^{\rm DSM}_{\pi}(\varepsilon)= 0$.

Second, for $\varepsilon=0$ but arbitrary $\alpha$, by using $\gamma_{n,\varepsilon=0}=-\pi/2$, 
the linear system \eqref{eq:linearsystem}, expressed in terms of the rescaled coefficients $\hat{c}^{\chi}_{n}=\sqrt{\kappa_{n,\varepsilon=0}}{c}^{\chi}_{n,\varepsilon=0}$, simplifies to
\begin{align}\nonumber
        0&=\hat{c}^{\chi}_{ 0}+\cos{\alpha}\,\hat{c}^{-\chi}_{0}+i\sin{\alpha}\,\hat{c}^{\chi}_{1},\\
        0&=2\hat{c}^{\chi }_{n}-i\sin{\alpha}\lb \hat{c}^{\chi}_{n+1}-\hat{c}^{\chi}_{n-1}\rb.
\end{align}
The physical solution of the recursion relation is (we here assume $\cos\alpha>0$)
\be
        \hat{c}^{+}_0=-\hat{c}^{-}_0,\quad\hat{c}^{\chi}_{ n}=\lb i\tan{\frac{\alpha}{2}}\rb^n\hat{c}^{\chi}_{0}.
\ee
Without need for a cut-off $N$ and up to normalization,
we thereby arrive at the exact solution 
\be        \label{eq:Dirac_zero}
        \ket{\Psi_{\varepsilon=0}}\propto\sum_{\chi=\pm}\chi\sum_{n\geq 0} \frac{1}{\sqrt{\kappa_{n,0}}}\lb i\tan{\frac{\alpha}{2}}\rb^n\ket{\psi^\chi_{n,0}}_{\boldsymbol{\sigma}}\ket{\chi}_{\boldsymbol{\tau}}.
\ee
Clearly, the phase shift is again given by $\theta^{\rm DSM}_\alpha(\varepsilon=0)=\pi$.
Remarkably, the superposition state \eqref{eq:Dirac_zero} involves evanescent contributions even though no   
surface state exists for $\varepsilon=0$ with $B=0$ and $k_0=0$, see Eqs.~\eqref{eq:dispersion_Dirac} and \eqref{eq:constraint_Dirac}.
An analogous calculation leads to $\theta^{\rm DSM}_{\pi-\alpha}(0)=0$.
    
For finite $\alpha$ and finite $\varepsilon$, we solve the problem numerically as described above.
As shown in Fig.~\ref{fig:3}, we then find a finite gauge invariant phase shift, which we define as
\be\label{dsmphaseshift}
\Phi^{\rm DSM}_\alpha(\varepsilon)=\theta^{\rm DSM}_\alpha(\varepsilon)-\theta^{\rm DSM}_\pi(\varepsilon).
\ee
(The reason for substracting the phase for $\alpha=\pi$ is explained in Sec.~\ref{semiclassics}.)
For small $\alpha$ and $\varepsilon$, this phase shift turns out to be small compared to the corresponding phase shifts
in WSMs, see Sec.~\ref{semiclassics}.  Since the main focus of this work in on the WSM case, we leave a detailed (semiclassical) discussion of phase shifts in DSMs to future studies. 

\section{Results and comparison to semiclassics}
\label{semiclassics}

The semiclassical theory for Fermi arcs in WSMs in a magnetic field is well established \cite{potter2014quantum,zhang2016quantum}.    According to this standard picture, fermions in the chiral LL tunnel into a Fermi-arc state upon reaching the surface.
The Lorentz force then drives the fermion along the arc to the other Weyl cone projection of opposite chirality, where it can tunnel back into the bulk and thereby escape from the surface.
In a slab geometry, this process is repeated on the opposite surface, and the semiclassical trajectory forms a closed Weyl orbit which can be described using semiclassical quantization \cite{potter2014quantum,zhang2016quantum}.   

In the half-space geometry, the semiclassical trajectory is open and no quantization is expected.    
This enables us to disentangle bulk and surface contributions. 
The latter are determined by the semiclassical equations of motion for an electron moving along the Fermi arc 
(with $\mb k=\mb k_\perp$) \cite{sundaram1999wave,potter2014quantum},
\be    \label{eq:semiclass_eom}
  \dot{\mb k}=-\frac{1}{\ell_B^2}\mb{v}_{\mb k}\times\hat{\mb{e}}_z,\quad \dot{\mb{r}}=\mb{v}_{\mb k}=\nabla_{\mb k} \varepsilon(\mb k),
\ee
where $\mb v_{\mb k}$ is the group velocity in the $x$-$y$ plane and $\varepsilon=\varepsilon(\mb k)$ is the arc  dispersion relation.    Here, we neglect the anomalous velocity contribution due to the Berry curvature of generic Fermi-arc states \cite{sundaram1999wave,wawrzik2021infinite}. 
This approximation can be justified by noting that the Berry curvature vanishes for a straight arc  
and we consider the small-$\alpha$ case below.  As a consequence, $\dot{\mb k}$ is tangential to the energy contour.

\subsection{Phase shifts accumulated along Fermi-arc curves}

\begin{figure}[t]
\centering
\includegraphics[width=0.9\columnwidth]{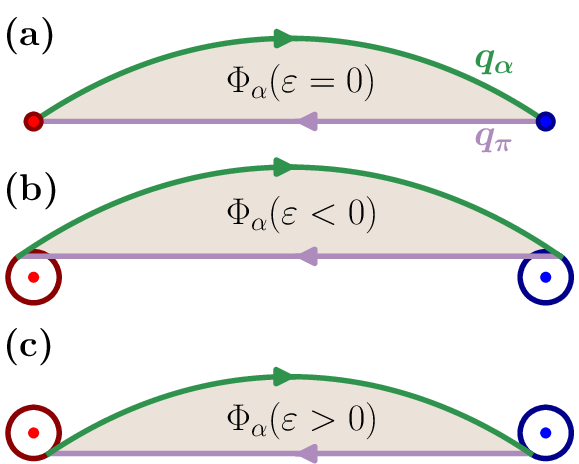}
 \caption{ Schematic illustration of closed trajectories in the surface momentum plane used for 
 the semiclassical calculation of the gauge invariant phase $\Phi_\alpha(\varepsilon)$ in Eq.~\eqref{semiphase}. 
 A curved Fermi arc (green) with $0<\cos\alpha<1$ is joined with a straight Fermi arc (purple)
 for $\alpha=\pi$. Arrows indicate the direction of $\dot{\mb k}$ as described by Eq.~\eqref{eq:semiclass_eom}.
 We show the corresponding
 closed trajectories (a) for zero energy ($\varepsilon=0$), (b) for $\varepsilon<0$, and (c) for $\varepsilon>0$.
 For $\varepsilon\ne 0$, the curved arc termination points, $(k_x,k_y)=(\pm k_{\varepsilon\alpha},q_\alpha(\varepsilon,k_{\varepsilon\alpha}))$, differ 
 from the Weyl node projections $(\pm k_0,0)$ corresponding to the circle centers in panels (b) and (c). 
 To match the arc termination points of the curved arc and the straight reference arc, we employ a rescaling $k_0\to 
  k_{\varepsilon\alpha}$ for the straight arc case.  For details, see main text. }
 \label{fig:4}
\end{figure}

We first consider the phase shift $\theta_\alpha(\varepsilon)$ between the chiral LLs in Eq.~\eqref{eq:phaseshift} for a curved Fermi arc with
 $0<\cos\alpha<1$.  In a semiclassical picture, this phase shift can be estimated by a phase-space integral of the schematic form 
\begin{equation}\label{theta1}
 \theta_\alpha(\varepsilon)=\int d\mb{r}\cdot\lb\mb k-\frac{e}{c}\mb{A}\rb.
\end{equation}
For gauge invariant phases, we need closed trajectories in real space. 
This issue is closely related to the fact that the quantum-mechanical phase shift $\theta_\alpha(\varepsilon)$ discussed in Sec.~\ref{sec3b} becomes gauge invariant only after switching to a phase shift difference.  For the semiclassical counterpart, we resolve this issue by introducing a straight reference arc which reconnects the termination points of the curved Fermi arc.  We thereby obtain a closed trajectory in the surface momentum plane, see Fig.~\ref{fig:4}, where the phase $\Phi_\alpha(\varepsilon)$ accumulated along the trajectory is gauge invariant.
To ensure that also the corresponding real-space trajectory is closed, we recall that the transformation 
$\alpha\to\alpha+\pi$ inverts the sign of the group velocity component $v_y$, and thus of $\dot{k}_x$,
see Eq.~\eqref{eq:semiclass_eom}.  In effect, this allows for a closed motion in the surface momentum plane,
where the straight reference arc is chosen to have $\alpha=\pi$. 

The above procedure is straightforwardly implemented at zero energy ($\varepsilon=0$), 
where the arc termination points are at $(k_x,k_y)=(\pm k_0,0)$ for all values of $\alpha$, see Fig.~\ref{fig:4}(a). 
On the quantum level, we then consider the phase shift difference $\Phi_\alpha(\varepsilon=0)=\theta_\alpha(0)-\theta_\pi(0)$, where $\theta_\pi(0)=\pi$, see Eq.~\eqref{pilock}. 

The situation becomes more intricate for $\varepsilon\ne 0$ since now the curved arc termination points, $(k_x,k_y)=(\pm k_{\varepsilon\alpha},q_\alpha(\varepsilon,k_{\varepsilon\alpha}))$, differ from the corresponding Weyl node projections at $(\pm k_0,0)$.
(We recall that $k_{\varepsilon\alpha}$ follows by solving Eq.~\eqref{eq:norm_constraint_main}, see also the estimate in Eq.~\eqref{eq:low_dispersion}.
Moreover, the function $q_\alpha(\varepsilon,k_x)$ parameterizing the Fermi-arc contour at energy $\varepsilon$ has been defined in Eq.~\eqref{eq:contour}.)
For the straight reference arc,  we therefore consider a system with rescaled Weyl node separation, $k_0\to k_{\varepsilon\alpha}$,
at energy $\varepsilon\to \bar\varepsilon=q_\alpha(\varepsilon,k_{\varepsilon\alpha})$. 
The arc termination points for the straight reference arc are then located at $(\pm k_{\varepsilon\alpha},\bar\varepsilon)$ and match the 
 termination points of the curved arc, see Fig.~\ref{fig:4}(b,c). We note that the energy of the reference arc differs from the energy of the curved Fermi arc. We can ensure only in this manner that both arc contours connect at their termination points and enclose a finite area in momentum space.
No need for such a construction would arise for Weyl orbits in a slab geometry, 
where tunneling processes via bulk states take care of the corresponding momentum shifts between arc termination points on opposite surfaces. 
The advantage of our approach is that bulk states do not appear explicitly in the semiclassical calculation.

\begin{figure*}[t]
\centering
\includegraphics[width=\textwidth]{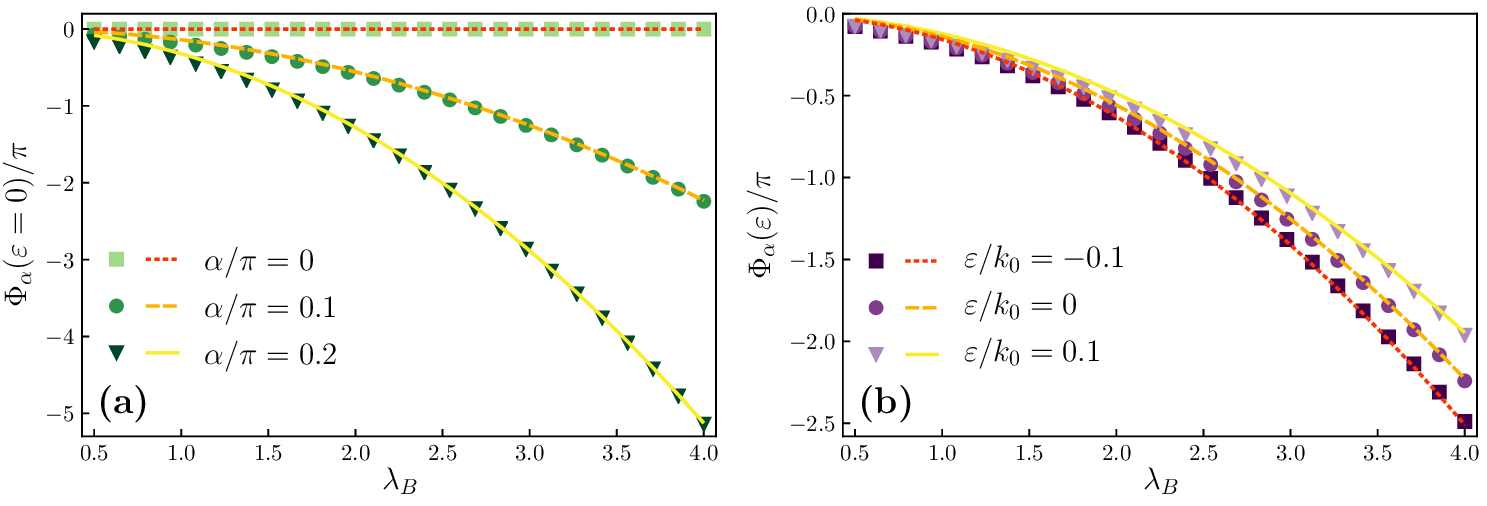}
\caption{ 
Comparison of quantum-mechanical results and semiclassical estimates for the gauge invariant Fermi-arc surface-state phase shift $\Phi_\alpha(\varepsilon)$, see Eqs.~\eqref{qmphase} and \eqref{semiphase2}, respectively.  The phase $\Phi_\alpha(\varepsilon)$ is shown as function of the parameter $\lambda_B=k_0\ell_B$ and gives the phase accumulated along the closed
trajectories in surface-momentum space illustrated in Fig.~\ref{fig:4}.
Symbols show the numerical solution of the quantum problem, and curves show the corresponding 
semiclassical predictions. Integer multiples of $2\pi$ have been added to obtain smooth curves.  (a) Zero-energy case ($\varepsilon=0$) for different $\alpha$.
(a) Case $\alpha/\pi=0.1$ for different energies $\varepsilon$.  }
\label{fig:5}
\end{figure*}

On the quantum level, we then define the gauge invariant phase shift difference as
\be\label{qmphase}
    \Phi_\alpha(\varepsilon)=\theta_\alpha(\varepsilon)-\bar{\theta}_\pi \left( -q_\alpha(\varepsilon,k_{\varepsilon\alpha})\right)-\Phi^{\rm DSM}_\alpha(\varepsilon),
\ee
where $\bar{\theta}_\pi$ follows by solving the linear system \eqref{eq:linearsystem} with
the rescaled parameter $\lambda_B\to \bar{\lambda}_B= \ell_B^2k_{\varepsilon\alpha}$.
For a comparison to semiclassical results, in Eq.~\eqref{qmphase}, we also subtract the phase shift difference $\Phi_\alpha^{\rm DSM}(\varepsilon)$, see 
Eq.~\eqref{dsmphaseshift}, for the corresponding DSM case as shown in Fig.~\ref{fig:3}.  

On the semiclassical level, the above gauge invariant phase shift takes the form  
\begin{align}\nonumber
    \Phi_\alpha(\varepsilon)&=\oint d\mb{r}\cdot\lb\mb{k}-\frac{e}{c}\mb{A}\rb\\
    &=-\ell_B^2\int_{-k_{\varepsilon\alpha}}^{k_{\varepsilon\alpha}} dk_x\,\lb q_\alpha(\varepsilon,k_x)-q_\alpha(\varepsilon,k_{\varepsilon\alpha})\rb, \label{semiphase}  
\end{align} 
As illustrated in Fig.~\ref{fig:4}, the phase $\Phi_\alpha(\varepsilon)$ in Eq.~\eqref{semiphase} corresponds to the
momentum-space area enclosed by the curved Fermi arc and the straight reference arc.  
Assuming $|\varepsilon|\ll k_0$, we find
\begin{align}\label{semiphase2}\nonumber
    \Phi_\alpha(\varepsilon)&\simeq \ell_B^2k_0^2\frac{2\alpha\cot{(2\alpha)}-1}{\sin{\alpha}}\\
    &+2\ell_B^2\varepsilon k_0 \lb 1+\alpha\tan{\alpha}-\frac{2\cos^2{\alpha}}{3-\cos{(2\alpha)}}\rb.
\end{align}
In Fig.~\ref{fig:5}, for small energies $\varepsilon$, we compare  quantum-mechanical results for $\Phi_\alpha(\varepsilon)$ obtained numerically from Eq.~\eqref{qmphase} to the corresponding semiclassical predictions in Eq.~\eqref{semiphase2}. 
We find quantitative agreement both for different arc curvatures, see Fig.~\ref{fig:5}(a), and for different energies, see Fig.~\ref{fig:5}(b). 
It is worth noting that the semiclassical description remains accurate even for large magnetic fields with $\lambda_B<1$.

\subsection{Fermi-arc lifetime and semiclassical traversal time}

 \begin{figure*}[t]
        \centering
        \includegraphics[width=\textwidth]{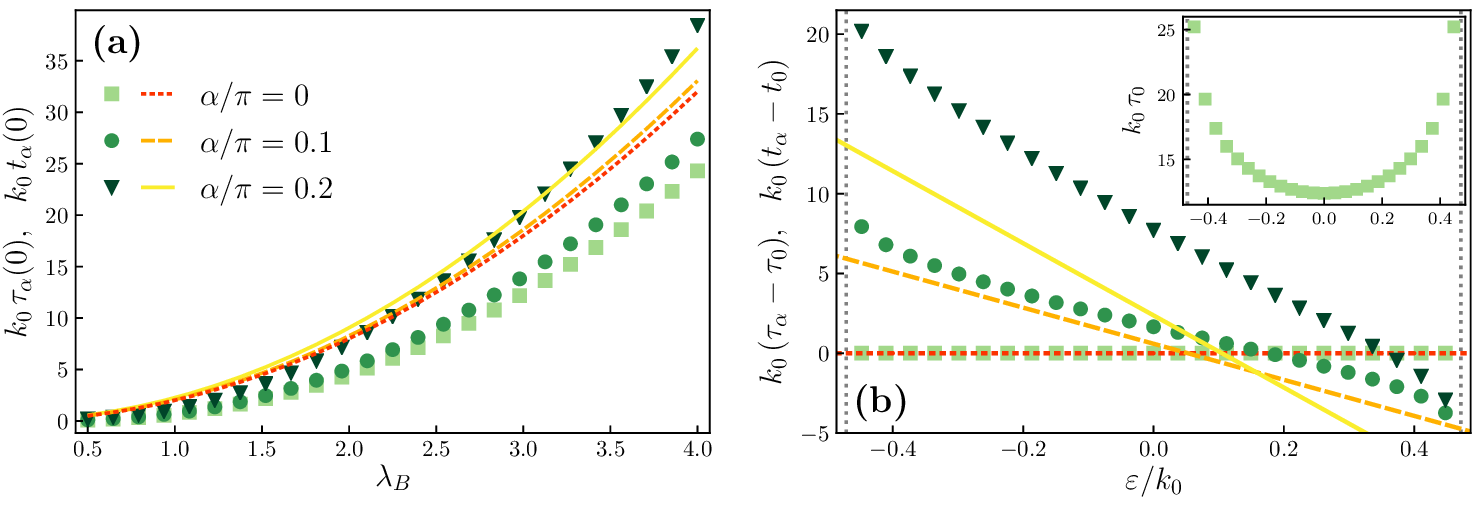}
        \caption{
        Quantum-mechanical results for the Fermi-arc lifetime $\tau_\alpha(\varepsilon)$ [symbols] 
        and for the semiclassical arc traversal time $t_\alpha(\varepsilon)$ [curves],
        see Eqs.~\eqref{eq:lifetime} and \eqref{eq:talpha}, respectively, for several values of $\alpha$ and given in units of $k_0^{-1}$.
        (a)  Dependence of $\tau_\alpha$ and $t_\alpha$ on $\lambda_B$ for $\varepsilon=0$. 
        (b) $\tau_\alpha-\tau_0$  vs $\varepsilon$ in the ultra-quantum regime (delimited by vertical dotted lines) 
        for $\lambda_B=3$. Different symbols are for different $\alpha$ as in panel (a). We also show the corresponding semiclassical traversal time differences $t_\alpha-t_0$. 
        The inset shows the straight-arc lifetime $\tau_{\alpha=0}$ vs $\varepsilon$, again for $\lambda_B=3$.
        \label{fig:6} }
\end{figure*}

As discussed in Sec.~\ref{sec1}, one expects that Fermi-arc surface states acquire a finite \emph{lifetime} $\tau_\alpha(\varepsilon)$ in a finite magnetic field $B\ne 0$.  
The lifetime describes the time scale for escaping into the bulk via the chiral LLs and  follows from the general relation \cite{Wigner1955,Smith1960}
\be\label{eq:lifetime}
        \tau_\alpha(\varepsilon)=\frac{d\theta_\alpha(\varepsilon)}{d\varepsilon}.
\ee
Indeed, as first shown in a seminal work by Wigner \cite{Wigner1955}, the energy derivative of the phase shift $\theta_\alpha(\varepsilon)$ encodes the time delay of a scattered particle which in turn is directly linked to its lifetime. 
We note that the phase shift \eqref{eq:lifetime} includes DSM contributions.
Being a physical observable, Eq.~\eqref{eq:lifetime} is gauge invariant.  We
compute Eq.~\eqref{eq:lifetime} numerically using the quantum-mechanical approach detailed in Sec.~\ref{Landau}.

On the semiclassical level, we define another time scale, namely the \emph{traversal time} $t_\alpha(\varepsilon)$.  This is the time required to traverse the Fermi arc from one termination point to the other. Since the lifetime is due to the escape of Fermi-arc electrons into the bulk at the arc termination points, one expects that $t_\alpha(\varepsilon)$ is of the same order   as $\tau_\alpha(\varepsilon)$. Even though these two time scales are not related to each other in a strict mathematical sense, one expects on physical grounds that they should exhibit similar behavior.  We therefore compare them in some detail below.
The semiclassical traversal time follows with Eq.~\eqref{eq:contour} in the gauge invariant form
\be \label{eq:arctime}
        t_\alpha(\varepsilon) =\ell_B^2\int_{-k_{\varepsilon\alpha}}^{k_{\varepsilon\alpha}}dk_x\,\sqrt{1+\lb\frac{\partial q_\alpha(\varepsilon,k_x)}{\partial k_x}\rb^2}|\, \mb{v}_{\mb{k}}|^{-1}.
\ee
Simple analytical expressions, cf.~Eqs.~\eqref{eq:contour} and \eqref{eq:low_dispersion}, follow for $|\alpha|\ll 1$ by expanding in $\alpha$ up to second order. We then obtain the semiclassical estimate
\be\label{eq:talpha}
    t_\alpha(\varepsilon)\simeq 2\ell_B^2\lsb k_0\lb 1+\frac{\alpha^2}{3}\rb-2\varepsilon\alpha\rsb.
\ee
We note that for a straight arc ($\alpha=0$), the energy-independent traversal time 
$t_0=2\ell_B^2k_0$ results.

In Fig.~\ref{fig:6}, we compare the semiclassical traversal time $t_\alpha(\varepsilon)$ to the quantum-mechanical lifetime $\tau_\alpha(\varepsilon)$.
As shown in Fig.~\ref{fig:6}(a), the zero-energy lifetime diverges with increasing $\lambda_B$ (i.e., with decreasing magnetic field), where the stable
$B=0$ Fermi arcs are approached.   The semiclassical traversal time qualitatively captures this behavior, but no
quantitative agreement between $t_\alpha(\varepsilon)$ and $\tau_\alpha(\varepsilon)$ is found.
As shown in the inset of Fig.~\ref{fig:6}(b), the lifetime of the straight arc ($\alpha=0$) increases with $|\varepsilon|$ and diverges upon reaching the $n=1$ bulk LL.  (We recall that our construction in Sec.~\ref{sec3b} is  limited to the ultra-quantum regime. For energies above the bulk gap of the $n=1$ LL, the $n=1$ LL contributes in terms of propagating states.)
Since the semiclassical estimate for $t_0$ is independent of energy,  the energy dependence of $\tau_\alpha(\varepsilon)$ shown
in the inset of Fig.~\ref{fig:6}(b) hints at quantum effects beyond semiclassics. 
    
To compare the two time scales $\tau_\alpha(\varepsilon)$ and $t_\alpha(\varepsilon)$ for curved arcs, we have substracted the respective
$\alpha=0$ contributions, and consider $\tau_\alpha-\tau_0$ and $t_\alpha-t_0$ in the main 
panel of Fig.~\ref{fig:6}(b).  We find that both quantities are approximately linear functions 
of energy (at low energies). The lifetime differences $\tau_\alpha(\varepsilon)-\tau_0(\varepsilon)$ are again qualitatively captured 
by the corresponding traversal-time differences $t_\alpha(\varepsilon)-t_0$ (up to a constant off-set). 

We conclude that while the Fermi-arc lifetime $\tau_\alpha(\varepsilon)$ includes 
quantum contributions beyond semiclassics, essential low-energy features are captured by the semiclassical traversal time, at least in a qualitative fashion.

\section{Discussion}\label{discussion}

In this work, we have studied the eigenstates of a four-band continuum model for a WSM in a half-space geometry, with a 
magnetic field perpendicular to the surface.  At low energies in the ultra-quantum regime dominated by the zeroth LL in the bulk, eigenstates are superpositions of in- and out-going chiral $n=0$ LL states coupled by  evanescent surface
states originating from $n\ne 0$ LL states.  The latter states replace the $B=0$ Fermi-arc surface state, which 
 acquires a finite lifetime for $B\ne 0$ and hence is not a stable solution.

We have compared our quantum-mechanical results with the corresponding semiclassical estimates by calculating the phase shift between in- and out-going $n=0$ chiral LL states with the corresponding semiclassical results.
These results depend on the energy $\varepsilon$ and on a boundary parameter $\alpha$ determining the Fermi-arc curvature for $B=0$.
According to Refs.~\cite{potter2014quantum,zhang2016quantum}, the coupling between the chiral LLs is established by a semiclassical motion of fermions along the arc due to the Lorentz force.  For the phase shifts, we find quantitative agreement between the quantum description and semiclassical estimates.  
Moreover, from the energy derivative of the phase shift, one can define the lifetime of the
Fermi-arc state.  By comparing the lifetime to the semiclassical arc traversal time, we have argued that quantum contributions beyond semiclassics are important for the lifetime. Our results indicate that quantum corrections remain significant upon lowering the magnetic field strength or when increasing the Fermi energy.
Understanding the lifetime of Fermi-arc surface states in an electromagnetic environment is a prerequisite for surface-sensitive tests such as 
quasi-particle interference experiments \cite{morali2019fermi}.
In the future, the theoretical modeling of such experiments could also profit from our explicit numerical construction of the eigenstates.

Our results are, at least qualitatively, consistent with numerical work on thin WSM films employing lattice models \cite{zhang2016quantum,abdulla2022time}, hybrid models \cite{benito2020surface}, and wave packet simulations \cite{yao2017simulation}.   The continuum approach used here employs a boundary condition which allows one to disentangle bulk and surface contributions to semiclassical trajectories.
Our analysis shows that a semiclassical phase-space integral along the Fermi arc provides  accurate estimates for phase shifts.  When extending our arguments
to a slab geometry or to thin films, one can describe the phase shift associated with Weyl orbits. This phase shift is observable in quantum magnetoconductance oscillations experiments, see Refs.~\cite{zhang2019ultrahigh,nair2020signatures,li2017evidence} for recent reports.  
Similar phase shifts are also expected to appear in transport experiments on WSM junctions 
with hetero-chiral Fermi arcs at the interface \cite{chaou2023magnetic}.
Our results justify semiclassical explanations of these experiments and provide analytical estimates for a minimal model that incorporates the 
Fermi-arc curvature. Importantly, the observability of quantum oscillations from Weyl orbits crucially depends on the comparison between the time needed to traverse the Fermi surface and the scattering time \cite{Bulmash2016}. Our estimates improve the evaluation of the former.

A more direct measurement of the traversal time can be devised along the lines of Ref.~\cite{baum2015current}. 
In the regime considered in our work, one can indeed consider a setup with two gates generating an electric field on one surface and measure the current on the opposite surface. As a consequence of the described hybridization of bulk and surface states, a pulsed electric field generates a current response on the opposite surface, within a duration given by the traversal time.

We have been able to make substantial progress, and in some cases even obtained exact analytical solutions, 
since the studied four-band WSM model has decoupled Weyl nodes in the bulk.  Omitting bulk Weyl-node coupling terms, e.g., a mass term $m \sigma^0\tau^x$, 
is typically justified for materials with well-separated Weyl nodes.  Indeed, assuming a Weyl node separation $2k_0\simeq 2\,\text{\AA}^{-1}$, Eq.~\eqref{lambdaB} gives $\lambda_B\simeq 2.57$ for $B\simeq 1\,\text{T}$.    The hybridization of LLs corresponding to different Weyl nodes is then exponentially suppressed by a factor $\mr{e}^{-\lambda_B^2}\simeq 0.0014$.  We conclude that
only for much smaller $k_0$  and/or stronger $B$, effects of bulk Weyl-node coupling are expected to become relevant.
For such cases, one expects a bulk gap for the hybridized $n=0$ LLs. As  a consequence, the chiral anomaly will eventually break down, and a  
non-monotonic magnetoconductance should appear \cite{chan2017emergence,saykin2018landau}. 
While such phenomena are not present in our study, they are unavoidable in lattice  models.  In fact, we believe that they obscure 
a semiclassical interpretation of previous numerical studies of WSM thin films \cite{zhang2016quantum,abdulla2022time,benito2020surface,yao2017simulation}.
Studying the effects of chiral mixing, e.g., by including the mass term $m\sigma^0\tau^x$ in our approach, 
is an interesting direction for future work.    Notably, numerical works in the Hofstadter regime suggest that depending on the exact nature of the Weyl node annihilation associated with the opening of the gap, the resulting insulating system can be either trivial or topological \cite{abdulla2022time,abdulla2024pairwise}.
In the latter case, localized topological surface states might emerge in the gap of the hybridized $n=0$ LLs. Such states seem to be outside the reach of the established semiclassical picture. 

The above-mentioned subtleties are absent if the magnetic field is oriented parallel to the axis along 
the Weyl node separation ($\hat {\mb e}_x$ in our case).
This scenario was studied for a thin-film geometry \cite{benito2020surface}, where a much smaller surface-bulk hybridization was reported, consistent with the semiclassical point of view.
Magnetic fields oriented in the surface plane generally result in qualitatively different physics \cite{tchoumakov2017magnetic,behrends2019landau} than reported here.

Our work has also covered the limiting DSM case. The considered Dirac Hamiltonian is 
an appropriate effective model as long as crystal symmetries protect the Dirac node degeneracy.
It would be interesting in a future study to apply our approach and
compare numerical and analytical solutions at $B\ne 0$ to the semiclassical description 
of topologically trivial surface states at $B=0$.
    
 In view of the recent experimental progress on magnetic WSMs \cite{bernevig2022progress}, such as Co$_2$MnGa \cite{belopolski2019discovery} and Co$_3$Sn$_2$S$_2$ \cite{liu2019magnetic,morali2019fermi},  we are optimistic that Weyl orbit physics will soon be clearly established also beyond DSMs and non-centrosymmetric crystals. 
The underlying physics of such compounds should be captured by our results.
    
\begin{acknowledgments} 
We thank M. Breitkreiz, P. Brouwer, A. Chaou and V. Dwivedi for discussions.
We acknowledge funding by the Deutsche Forschungsgemeinschaft (DFG, German Research Foundation) under Projektnummer 277101999 - TRR 183 (project A02) and under Germany's Excellence Strategy - Cluster of Excellence Matter and Light for Quantum Computing (ML4Q) EXC 2004/1 - 390534769.
The data underlying the figures in this work can be found at the zenodo site:  \href{https://doi.org/10.5281/zenodo.14062003}{https://doi.org/10.5281/zenodo.14062003}.
FB acknowledges financial support from the TOPMASQ Project, CUP E13C24001560001, Spoke 5 of the National Quantum Science and Technology Institute (NQSTI), PE0000023 of the Piano Nazionale di Ripresa e Resilienza (PNRR), financed by the European Union - NextGenerationEU.
\end{acknowledgments}


\appendix
\section{Surface state solutions}\label{surface_states}

Here we provide detailed derivations for the $B=0$ surface states given in Sec.~\ref{sec2c}.
We begin with the topologically trivial surface states for the \emph{Dirac semimetal} case, $k_0= 0$,
described by $H=\mb{k}\cdot\boldsymbol{\sigma}\tau^z$.   After the unitary transformation
$U_\alpha=\exp{\lb\frac{i}{2}\alpha\sigma^y\tau^x\rb}$, we  obtain
\begin{align}\nonumber
        \tilde{H}_\alpha=U_\alpha HU_\alpha^\dagger&=\lb k_x\sigma^x+k_y\sigma^y\cos{\alpha}-i\partial_z\sigma^z\rb\tau^z\\
        &+k_y\sigma^0\tau^y\sin{\alpha}.
\end{align}
This transformation is convenient since it eliminates the boundary parameter $\alpha$ from the boundary condition, $U_\alpha M_\alpha U_\alpha^\dagger=M_0=\sigma^z\tau^x$, see Eq.~\eqref{eq:Malpha}.
Note that the unitary transformation leaves the current operator $j^z$ invariant.  Therefore,  eigenstates $\ket{\tilde{\Psi}}$ with $\tilde{H}\ket{\tilde{\Psi}}=\varepsilon\ket{\tilde{\Psi}}$ 
must satisfy the boundary condition $M_0\tilde{\Psi}(z=0)=\tilde{\Psi}(z=0).$
We next make a (normalized) \emph{Ansatz} for a surface state confined to the half-space region $z\ge 0$,
\be\label{ansatz1}
\ket{\Psi}=\sqrt{\frac{\kappa}{2}}\bmat \tilde{\psi}^{+}\\\tilde{\psi}^{-}\emat_{\boldsymbol{\tau}},\quad \tilde{\psi}^{\chi}(z)=\mr{e}^{-\kappa z}\bmat 1\\\chi \mathrm{e}^{i\delta}\emat_{\boldsymbol{\sigma}},
\ee
where $\delta$ and $\kappa$ are a phase and an inverse penetration length, respectively. These quantities have yet to be determined, where Eq.~\eqref{ansatz1} satisfies the boundary condition for arbitrary $\delta$.
To construct energy eigenstates, we first note that the chiral components satisfy $M^{\boldsymbol{\sigma}}_\delta\ket{\tilde\psi^{\chi}}_{\boldsymbol{\sigma}}=
\chi\ket{\tilde\psi^{\chi}}_{\boldsymbol{\sigma}}$ for $M^{\boldsymbol{\sigma}}_\delta$ 
in Eq.~\eqref{eq:Mtausigma}. Eigenstates of $\tilde H$ thus obey  
\be
       \left( k_x\sigma^x+k_y\sigma^y\cos{\alpha}\right)\ket{\Psi}=\varepsilon M^{\boldsymbol{\sigma}}_\delta \ket{\Psi}.
\ee

For given in-plane momentum $\mb{k}_\perp$, the phase $\delta=\delta_\pm(\mb{k}_\perp)$ then follows from 
\be
\cos{\delta_\pm(\mb{k}_\perp)}=\frac{k_x}{\varepsilon_\pm(\mb{k}_\perp)},\quad \sin{\delta_\pm(\mb{k}_\perp)}=\frac{k_y\cos{\alpha}}{\varepsilon_\pm(\mb{k}_\perp)},
    \ee
with $\varepsilon_\pm(\mb{k}_\perp)$ in Eq.~\eqref{eq:dispersion_Dirac}.
    Inserting the corresponding Ansatz into the eigenproblem of $\tilde{H}$ confirms that $\varepsilon_\pm(\mb{k}_\perp)$ is the energy dispersion of the surface state
    and yields the inverse decay length $\kappa$ in the form
    \be
        \kappa=k_y\sin{\alpha}.
    \ee
    The normalization condition $\kappa>0$  implies  Eq.~\eqref{eq:constraint_Dirac} for physical  solutions.
     
Next, we construct the solution for a \emph{straight Fermi arc}, corresponding to the choice $\alpha=0$. For the purpose of generality, we here allow for a free parameter in the parameterization \eqref{eq:M_general}.
The trivial dependence of our results on this parameter (see below) helps to develop physical insight.  We consider the boundary condition \eqref{eq:BCM} with the matrix 
    \be
        M_{\gamma}'=M_{0,0,\gamma,0}=\sigma^z\lb \tau^x\cos{\gamma}+\tau^y\sin{\gamma}\rb.
    \ee
 The following results for $\gamma=0$ describe the $\alpha=0$ results in Sec.~\ref{sec2c} since $M'_{\gamma=0}=M_{\alpha=0}$ with $M_\alpha$ in Eq.~\eqref{eq:Malpha}.  (Note that $\delta$ in Eq.~\eqref{eq:M_general} is redundant for $\alpha=0$.) 
We choose the normalized Ansatz  
\be
    |\Psi\rangle=\sqrt{\frac{\kappa_+\kappa_-}{\kappa_++\kappa_-}}\bmat \psi^{+}\\\mr{e}^{i\gamma}\psi^{-}\emat_{\boldsymbol{\tau}},
\ee
with the chiral spinor components
\be  \label{eq:straight_arc_sol}
\psi^{\chi}(z)=\mr{e}^{-\kappa_\chi z}\bmat 1\\-i\chi \emat_{\boldsymbol{\sigma}}.
\ee
This \emph{Ansatz} satisfies the boundary condition.   From $H|\Psi\rangle=\varepsilon|\Psi\rangle$, we find
\be
        \varepsilon(k_y)=-k_y,\quad \kappa_\chi(k_x)=k_0+\chi k_x.
\ee
The normalization conditions $\kappa_+>0$ and $\kappa_->0$ for surface-state solutions restrict the 
in-plane momentum $k_x$ to the open interval $-k_0<k_x<k_0$. We thus obtain a physical Fermi arc for a model 
with two decoupled Weyl nodes in the bulk.  Here it turns out that the energy dispersion $\varepsilon(k_y)$ and the
inverse penetration length scales $\kappa_\chi$ are independent of $\gamma$.    This is expected since
the parametric freedom in the boundary condition increases with the number of higher-energy bands.
However, in this instance, we can extend the relation between the arc curvature and the corresponding boundary matrix parameterization further.
To this end, we note that a straight arc is characterized by a chiral dispersion along $\hat{\mb e}_y$, 
and consequently a maximal current flows along this direction.   
Accordingly, the found solutions are eigenstates of the in-plane current $j^y=\sigma^y\tau^z$, which is only possible since $M'_\gamma$ commutes with $j^y$.    We can therefore infer the necessary condition that a straight arc corresponds to a parameterization of $M$ with $[M,j^y]=0$. Note that for the parameterization $M_\alpha$ in Eq.~\eqref{eq:Malpha}, this condition is only met for $\alpha=0$~mod~$\pi$.  

In fact, we find \emph{curved Fermi arcs} for all other values of $\alpha$.
For solving the surface-state problem, we here use a different approach which  applies to a large family of parameterizations.
We first consider the eigenproblem $H_\chi \ket{\psi^{\chi}}_{\boldsymbol{\sigma}}=\varepsilon\ket{\psi^{\chi}}_{\boldsymbol{\sigma}}$ for a single Weyl node with chirality $\chi=\pm 1$, described by $H_\chi=\chi \mb k\cdot\boldsymbol{\sigma}+k_0\sigma^x$.    The most general evanescent and normalized solution at given energy $\varepsilon$ and in-plane momentum $\mb k_\perp$ is
    \be        \label{eq:evanescent_sol}
        \psi_{\varepsilon\mb k_\perp}^{\chi}(z)=\sqrt{\frac{\kappa_\chi}{\varepsilon^2+\kappa_\chi^2}}\mr{e}^{-\kappa_\chi z}\bmat \chi \varepsilon+i\kappa_\chi\\ k_x+\chi k_0+ik_y\emat_{\boldsymbol{\sigma}},
    \ee
where $\kappa_\chi(\varepsilon,\mb k_\perp)=\sqrt{(k_x+\chi k_0)^2+k_y^2-\varepsilon^2}$ is the inverse length scale
describing the decay of the surface state into the bulk.
The requirement that $\kappa$ is real restricts the energy of physical solutions to
\be        \label{eq:constraint_curved}
\varepsilon^2<\lb|k_x|-k_0\rb^2+k_y^2.
\ee
The solution with energy $\varepsilon$ in this interval is given by
\be        \label{eq:Ansatz_curved}
    \ket{\Psi_{\varepsilon\mb{k}_\perp}}=\bmat c_+\psi^{+}_{\varepsilon\mb{k}_\perp}\\c_-\psi^{-}_{\varepsilon\mb{k}_\perp}\emat_{\boldsymbol{\tau}},
\ee
    where $c_\pm$ are complex coefficients.
We now consider the boundary condition \eqref{eq:BCM} with a general Hermitian parameterization,
\be        
 M=\bmat X&Y\\Y^\dagger &Z\emat,\quad X=X^\dagger,\quad Z=Z^\dagger.
\ee
Here, we assume that $Y$ is invertible, which implies the condition $[M,\sigma^0\tau^z]\neq 0$ for a physical Fermi arc, see Sec.~\ref{sec2b}.  
Together with $M^2=\mathbbm{1}$, we obtain the identities $Z=-Y^{-1}XY$ and  $Y^{-1}X^2=Y^{-1}-Y^{\dagger}$.  It is then sufficient to consider the
upper two spinor components in the boundary condition $(\mathbbm{1}-M)\Psi_{\varepsilon\mb{k}_\perp}(z=0)=0$, since the lower two 
components are implied.  One can express the upper two components as a linear system of equations, $\mathbbm{B}\mathbf{c}=\mathbf{0}$, where
\be
        \mathbbm{B}(\varepsilon,\mb k_\perp)=\bmat (\sigma^0-X)\psi_{\varepsilon\mb{k}_\perp}^{+}(0) \quad -Y\psi_{\varepsilon\mb{k}_\perp}^{-}(0)\emat
\ee
is a $2\times 2$ matrix and $\mathbf{c}=\lb c_+,c_-\rb^T$ contains the coefficients in Eq.~\eqref{eq:Ansatz_curved}.
For  $M_\alpha$ in Eq.~\eqref{eq:Malpha}, we have $X_\alpha=\sigma^x\sin{\alpha}$ and $Y_\alpha=\sigma^z\cos{\alpha}$. 
Solutions of the boundary condition thus satisfy $\det(\mathbbm{B})=0$.
We then obtain a secular equation that gives analytical solutions for $k_y=q_\alpha(\varepsilon,k_x)$, where Eq.~\eqref{eq:contour} is the only solution satisfying Eq.~\eqref{eq:constraint_curved}.

We note that surface-state solutions for the opposite half-space ($z\leq 0$) with the same boundary condition~\eqref{eq:BCM} follow from the transformation $\alpha\rightarrow\pi-\alpha$. This is
because the transformation $\kappa_\chi\to-\kappa_\chi<0$, necessary for constructing physical states in this geometry, amounts to $\psi^\chi_{\varepsilon,k_x,k_y}(0)\to\lb \psi^\chi_{\varepsilon,k_x,-k_y}(0)\rb^*$. The corresponding secular equation $\det\lsb \mathbbm{B}^*(\varepsilon,k_x,-k_y)\rsb=0$ then yields $k_y=-q_\alpha=q_{\pi-\alpha}$.
 
\section{Surface spin polarization and current}
\label{spinArc}

    \begin{figure}[t]
        \centering
        \includegraphics[width=\columnwidth]{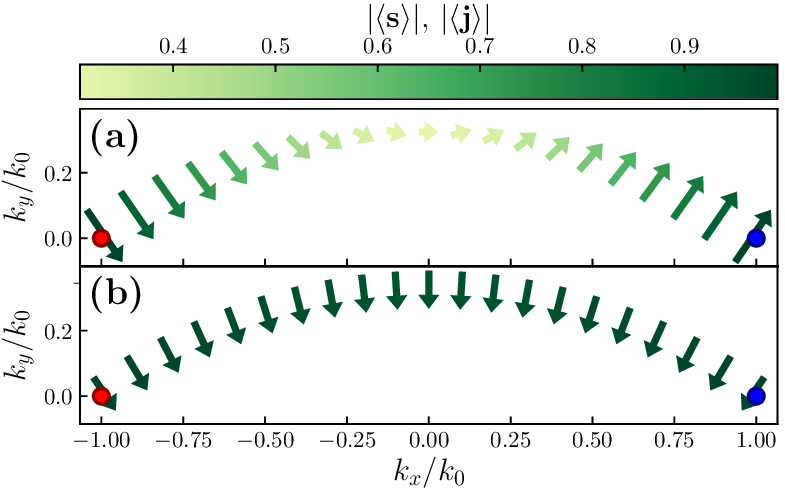}
        \caption{In-plane spin polarization and particle current corresponding to the Fermi-arc surface state for $B=0$, $\alpha/\pi=0.1$, and $\varepsilon=0$, shown
        as color-scale plots in the $k_x$-$k_y$ plane, cf.~App.~\ref{spinArc}.  Red and blue dots indicate the surface projections of the Weyl nodes, arrows show the in-plane components of the spin polarization $\braket{\mb s}$ [panel (a)] and of the particle current $\braket{\mb j}$ [panel (b)].}
        \label{fig:7}
    \end{figure}
    
\begin{figure}[t]
\centering
\includegraphics[width=\columnwidth]{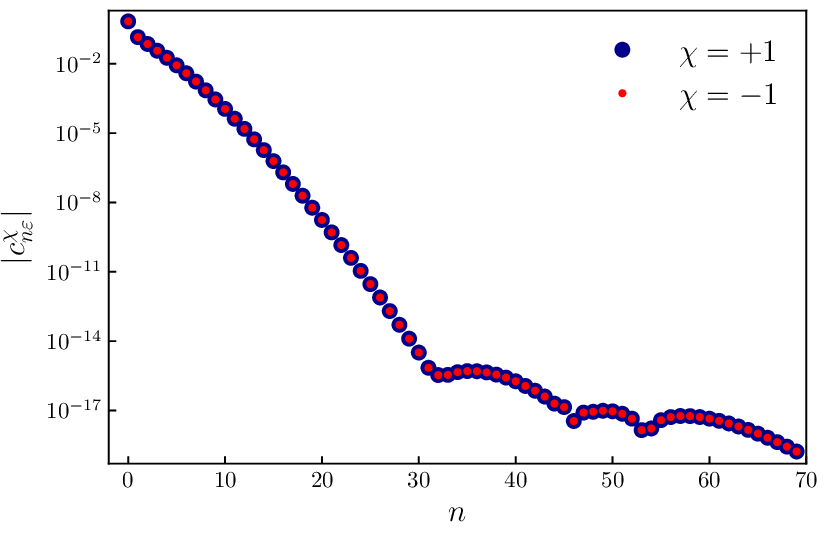}
\caption{Absolute value of the coefficients $c^\chi_{n\varepsilon}$ appearing in the superposition \eqref{eq:superposition}
vs order $n$.  Note the logarithmic scale for the coefficients.  The shown results were obtained by numerically solving 
Eq.~\eqref{eq:linearsystem} for $\varepsilon/k_0=0.1$, $\alpha/\pi=0.1$, $\lambda_B=1$, with a cut-off value of $N=70$.}
\label{fig:8}
\end{figure}

    \label{numerics}
    \begin{figure*}[t]
        \centering
        \includegraphics[width=\textwidth]{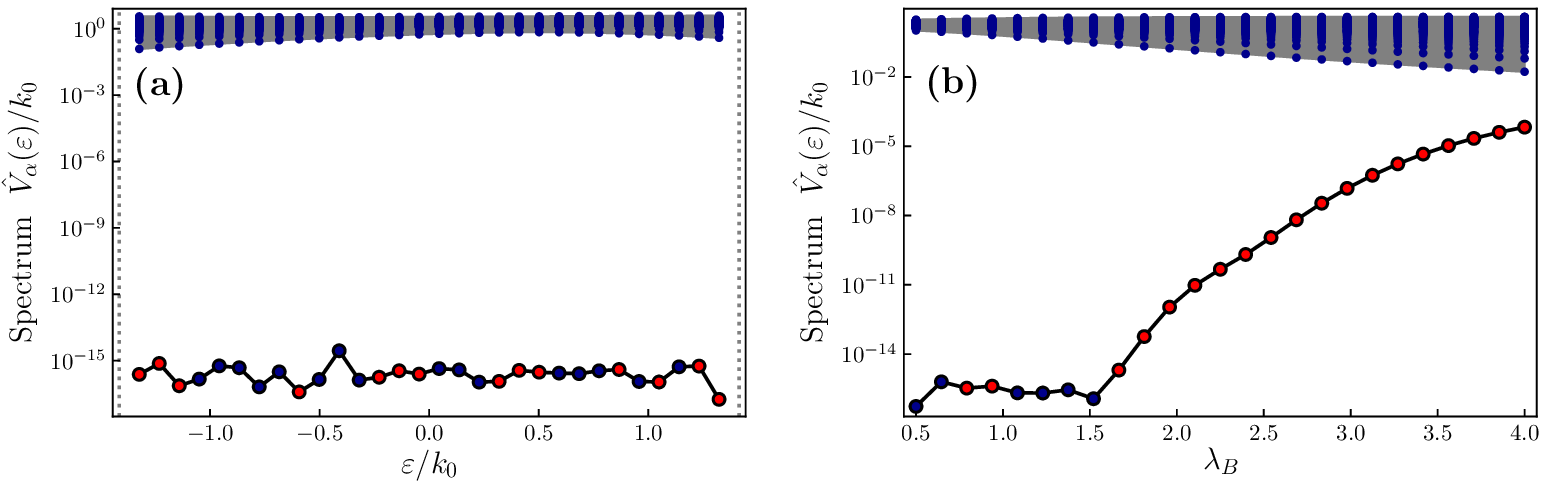}
        \caption{Spectrum of $\hat{V}_\alpha$ as obtained from Eq.~\eqref{eq:rescaled_V} for $\alpha/\pi=0.1$, with the cut-off value $N=100$. Note the semi-logarithmic scales.
        Blue (red) dots correspond to positive (negative) eigenvalues. Since $\mathbbm{V}_\alpha$ is positive semi-definite, negative eigenvalues indicate numerical errors. (a) Spectrum vs energy $\varepsilon$ (in units of $k_0$) in the ultra-quantum regime $|\varepsilon|<\sqrt 2/\ell_B$ delimited by the vertical gray line, for $\lambda_B= 1$.         (b) Spectrum vs $\lambda_B$ for $\varepsilon/k_0=0.1$.}
        \label{fig:9}
    \end{figure*}

    \begin{figure*}
        \centering
        {\includegraphics[width=\textwidth]{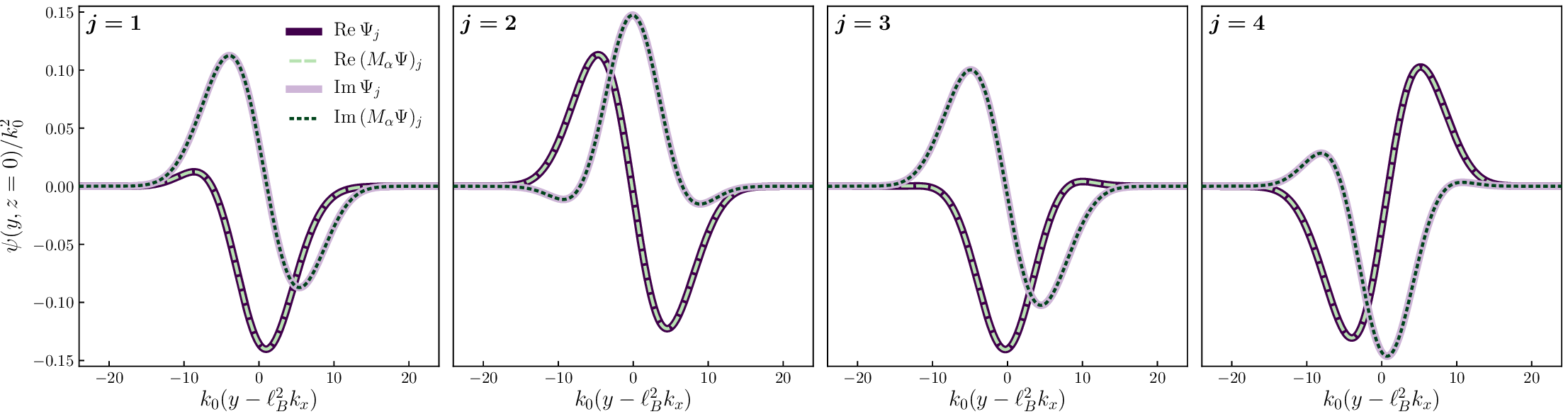}}
        \caption{Spinor components $\Psi_{j=1,2,3,4}$ of the eigenstate solution at the surface, $\Psi(y,z=0)=\lb\Psi_1(y),\Psi_2(y),\Psi_3(y),\Psi_4(y)\rb^T$, compared with the corresponding components of $M_\alpha\Psi(y,z=0)$. Parameters are given by $\varepsilon/k_0=0.1$, $\alpha/\pi=0.1$ and $\lambda_B=3$. We separately show the real and imaginary parts, which verify that the boundary condition $M_\alpha\Psi(y,z=0)=\Psi(y,z=0)$ is numerically satisfied to high accuracy.}
        \label{fig:10}
    \end{figure*}

In this Appendix, we discuss the spin texture related to $s^\mu=\sigma^\mu\tau^0$ (with $\mu=x,y,z$) along a Fermi arc for the $B=0$ case. 
Getting access to this type of quantity is an advantage of the four-band model with respect to two-band models \cite{bovenzi2018twisted,burrello2019field,Buccheri2022a}.
Furthermore, we compute the in-plane current $j^\mu=\sigma^\mu\tau^z$ generated by the Fermi arc. 
Given a normalized Fermi-arc solution $\ket{\Psi}$, we need to evaluate expectation values of the form
$\braket{\sigma^\nu\tau^\mu}=\int^\infty_0  dz\,\Psi^\dagger(z)\sigma^\nu\tau^\mu\Psi(z).$

For a straight arc ($\alpha=0$), the corresponding solutions in Eq.~\eqref{eq:straight_arc_sol} satisfy $\sigma^y\ket{\psi^{\chi}}_{\boldsymbol{\sigma}}=-\chi\ket{\psi^{\chi}}_{\boldsymbol{\sigma}}$, implying 
$\braket{\mb s}=k_x\hat{\mb e}_y/k_0$ and $\braket{\mb j}=-\hat{\mb e}_y$.

For curved arcs with $\alpha>0$, we use the general solution \eqref{eq:evanescent_sol} and perform the integration. The spin polarization follows from     
\be
        \braket{s^\mu}=\sum_{\chi=\pm}\frac{|c_\chi|^2}{2\kappa_\chi} \psi^{\chi\,\dagger}(0) \sigma^\mu\psi^{\chi}(0),
\ee
where the expression for the in-plane current only differs by a relative sign in the sum,
\be
        \braket{j^\mu}=\sum_{\chi=\pm}\chi\frac{|c_\chi|^2}{2\kappa_\chi} \psi^{\chi\,\dagger} (0)\sigma^\mu\psi^\chi(0).
\ee
Above, we have suppressed the  momentum dependence of $c_\chi$ and of $\ket{\psi^{\chi}}_{\boldsymbol{\sigma}}$.

Results obtained from the above expressions are shown in Fig.~\ref{fig:7}. Our model correctly reproduces the main features of Fermi arcs as experimentally detected. First and foremost, the chiral transport is shown by the current in Fig.~\ref{fig:7}(b). In addition, the spin polarization rotates along the arc as dictated by the fact that the spin orientation at the two termination points corresponds to the chirality of the Weyl nodes. 
This behavior is manifest in Fig.~\ref{fig:7}(a) and in accordance with the spin texture observed 
experimentally by spin-filtered angle-resolved photoemission spectroscopy \cite{Lv2015spin,Xu2016spin}.

\section{Numerical implementation of boundary conditions} \label{appnumerics}

In this Appendix, we discuss the numerical approach introduced in Sec.~\ref{Landau}.
Figure~\ref{fig:8} shows representative results for the coefficients ${\mb c}_\varepsilon$ 
in Eq.~\eqref{eq:superposition}, which are obtained by numerically solving Eq.~\eqref{eq:linearsystem} for a Landau level cut-off $N=70$.
These results already indicate that the numerical scheme is well controlled and convergent.
A non-trivial benchmark that is passed accurately by our numerical scheme is provided by the analytical solutions \eqref{eq:Dirac_straight} and \eqref{eq:Dirac_zero} for a DSM with $\alpha=0$ or $\varepsilon=0$, respectively.    

Let us next give additional details about our numerical approach.
To avoid numerical overflow (or underflow) when computing the matrix elements \eqref{eq:matrixelements} for a large cut-off $N$, it is convenient to compute the overlaps \eqref{eq:overlaps} using the recursion relation ($n\geq m>1$)
    \begin{align}\nonumber
        \braket{\varphi^{-}_{n}|\varphi^{+}_{m}}&=\frac{1}{\sqrt{nm}}\lb n+m-1-2\lambda_B^2\rb\braket{\varphi^{-}_{n-1}|\varphi^{+}_{m-1}}\\
        &-\sqrt{\frac{(n-1)(m-1)}{nm}}\braket{\varphi^{-}_{n-2}|\varphi^{+}_{m-2}},
    \end{align}
    with 
    \be
        \braket{\varphi^{-}_{n-m+1}|\varphi^{+}_{1}}=\frac{n-m+1-2\lambda_B^2}{\sqrt{n-m+1}}\braket{\varphi^{-}_{n-m}|\varphi^{+}_{0}}
    \ee
    and
    \be
        \braket{\varphi^{-}_{n-m}|\varphi^{+}_{0}}=\sqrt{\frac{2^{n-m}}{(n-m)!}}\lambda_B^{n-m}\,\mr{e}^{-\lambda_B^2}.
    \ee
    With these relations, we can easily employ a LL number cut-off of order $N=250$ or even larger.
    For all results shown in this work, we have carefully checked that results do not change when further increasing the cut-off.
    
    Numerical solutions are then found from the kernel of $V_\alpha(\varepsilon)$, i.e., from
    the matrix representation of $\mathbbm{V}_\alpha$ in the subspace with fixed $\varepsilon$ and $k_x$.
    Note that we physically expect a single solution in this subspace in the ultra-quantum regime.
    Consequently, the spectrum of $V_\alpha(\varepsilon)$ should have a single zero eigenvalue which is well separated from all other eigenvalues.
    Fig.~\ref{fig:9} shows representative results for the spectrum of the rescaled matrix $\hat{V}_\alpha$ obtained from Eq.~\eqref{eq:rescaled_V}.
    We find a non-degenerate, well-separated and vanishing eigenvalue for all $\varepsilon$ in the ultra-quantum regime.
    However, for $\lambda_B\agt 1.5$ (weak magnetic fields), numerical errors become slightly larger.  Nonetheless, our numerical solutions still satisfy the boundary condition as demonstrated for $\lambda_B=3$ in Fig.~\ref{fig:10}, 
    where we show the four components of the real and imaginary parts of $\Psi(y,z=0)$ and $M_\alpha\Psi(y,z=0)$, respectively.  The boundary condition $\Psi(y,z=0)=M_\alpha\Psi(y,z=0)$ is indeed satisfied to high precision for all values of $y$.

    \section{Shift of the reflected electron} \label{GHshift}

    The electronic Goos-H\"anchen effect is a quantum phenomenon, best described as a lateral shift of a 
    wave packet after reflection from a surface \cite{Chen2013gefluegel}. We can see an analogue of this effect in the system at hand at the level of the expectation value of the position operator. In particular, we can read Eq.~\eqref{eq:superposition} in the ultra-quantum regime $0<\varepsilon<\sqrt{2}/\ell_{B}$ as the superposition of an incoming wave in the chiral LL with $\chi=+1$, an outgoing wave in the chiral LL with $\chi=-1$ and a series of bound states, see Eq.~\eqref{eq:LLchiral}. Considering first a single momentum component $k_{x}$, the expectation value of the $y$-coordinate for an incoming electron arriving on the surface ($z=0$) is computed from the fundamental eigenmode of the harmonic oscillator in Eq.~\eqref{harmonicmodes} as $\left\langle y\right\rangle_{\mr{in}}=\ell_{B}^{2}\left(k_{x}+k_{0}\right)$. The momentum $k_{x}$ is conserved in the reflection process, and one readily sees that the electron leaving the surface has the expectation value 
    $\left\langle y\right\rangle_{\mr{out}}=\ell_{B}^{2}\left(k_{x}-k_{0}\right)$. 
    We note that the shift $\langle y\rangle_{\mr{out}}-\langle y\rangle_{\mr{in}}$ is gauge invariant.
    
    Following Ref.~\cite{Chen2013gefluegel}, we now generalize this argument and write an electronic wave packet formed by a superposition of plane waves with various momenta $k_{x}$ and, for simplicity, a Gaussian envelope function centered around momentum $p$ with spread $\Delta k$. Such a wavepacket, in the $\chi=+1$ block, has the form 
    \begin{equation}
            \Psi_{\varepsilon,\rm{in}}(\mb r)=\frac{1}{\sqrt{2\pi}}\int dk_{x}\,F\left(k_{x}-p\right)\mr{e}^{ik_{x}x}\psi_{0,\varepsilon}^{+}(y,z),
    \end{equation}
    with    $F\left(k_{x}\right)=\lb\sqrt{\pi}\Delta k\rb^{-\frac{1}{2}}\mr{e}^{-\frac{1}{2}\lb k_x/\Delta k\rb^2}$.
    The expectation value of the $y$-coordinate for the wavepacket arriving on the surface then follows as
    \begin{align}
        \left\langle y\right\rangle_{{\rm in}} =\ell_{B}^{2}\left(p+k_{0}\right).
    \end{align}
    Repeating the calculation for the outgoing wavepacket in the $\chi=-1$ block, 
    \begin{equation}
            \Psi_{\varepsilon,\mr{out}}(\mb r)=\frac{1}{\sqrt{2\pi}}\int dk_{x}\,F\left(k_{x}-p\right)e^{ik_{x}x}\psi_{0,\varepsilon}^{-}(y,z) ,
    \end{equation}
one finds $\left\langle y\right\rangle_{{\rm out}}=\ell_{B}^{2}\left(p-k_{0}\right)$. We conclude 
that the electronic Goos-H\"anchen shift is given by Eq.~\eqref{GHdeltay}. As this result is separately valid for each $k_x$, we expect it to hold for every choice of the envelope function.

\bibliography{references}

\end{document}